\documentclass[aps,physrev,reprint,groupedaddress]{revtex4-2}

\usepackage{graphics}
\usepackage{graphicx}
\usepackage{amsmath}

\begin{document}


\title{Neutron Beam Shaping by Ghost Projection}


\author{Andrew M. Kingston}
\email[]{andrew.kingston@anu.edu.au}
\thanks{corresponding author}
\altaffiliation{National Laboratory for X-ray Micro Computed Tomography (CTLab), Advanced Imaging Precinct, The Australian National University, Canberra, ACT 2600, Australia}
\affiliation{Department of Materials Physics, Research School of Physics, The Australian National University, Canberra, ACT 2600, Australia}

\author{Alaleh Aminzadeh}
\affiliation{Department of Materials Physics, Research School of Physics, The Australian National University, Canberra, ACT 2600, Australia}

\author{Lindon Roberts}
\affiliation{School of Mathematics and Statistics, Carslaw Building, University of Sydney, Camperdown NSW 2006, Australia}

\author{Jeremy M.~C.~Brown}
\affiliation{Optical Sciences Centre, Department of Physics and Astronomy, School of Science, Computing and Engineering Technologies, Swinburne University of Technology, Hawthorn, VIC 3122, Australia}

\author{Filomena Salvemini}
\affiliation{Australian Centre for Neutron Scattering, Australian Nuclear Science and Technology Organisation, Lucas Heights, NSW 2234, Australia}

\author{Joseph J.~Bevitt}
\affiliation{Australian Centre for Neutron Scattering, Australian Nuclear Science and Technology Organisation, Lucas Heights, NSW 2234, Australia}

\author{Ulf Garbe}
\affiliation{Australian Centre for Neutron Scattering, Australian Nuclear Science and Technology Organisation, Lucas Heights, NSW 2234, Australia}

\author{David M.~Paganin}
\affiliation{School of Physics and Astronomy, Monash University, Clayton, VIC 3800, Australia}


\date{\today}

\begin{abstract}
We present a method to shape a neutron beam and project any specified target image using a single universal patterned mask that is transversely displaced. The method relies on ``ghost projection'', which is a reversed form of classical ghost imaging. A set of sub-mask regions that combine to construct the required beam shape is computed; illumination of each region with the determined exposure time projects the shaped beam. We demonstrate this method experimentally, using the Dingo neutron imaging beamline at the OPAL nuclear research reactor (Australia). The ability to shape a neutron beam ``on demand'' allows selective dose delivery away from sensitive areas of samples, such as in cultural heritage artifacts. It also benefits irradiation techniques, e.g., in testing resilience of electronic components for space and defense technologies or neutron therapies.
\end{abstract}


\maketitle

\section{Introduction}
\label{sec:intro}



Controlled illumination and dose delivery of ionising radiation is at the core of a number of day-to-day techniques and technologies used in medical physics (imaging \cite{behling2018diagnostic, greene2017linear} and radiotherapy \cite{degiovanni2015history}), electronics production (semiconductor material doping \cite{crowder2013ion}), and testing or inspection of devices and materials critical to everyday life (medical devices \cite{wang2020non}, homeland security \cite{wells2012review,akcay2022towards}, space and defense industries \cite{hanke2008x,towsyfyan2020successes}, etc.). The bulk of these technologies utilize x rays, $\gamma$ rays, and charged particles due to their strong electromagnetic interaction with matter, and electromagnetic fields enabling precise control of their incident direction and dose deposition within objects and materials. However, until now, this has not been the case for neutron sources, which has led to a number of limitations of their applications in the same manner. The ability to arbitrarily shape neutron beams would enable accurate illumination of patterns onto sample surfaces and volumetrically tunable autoradiography, (i.e.,~selective dose delivery away from sensitive areas). This new functionality could be immediately leveraged to (i) improve our understanding of how neutrons kill cells in charged particle and neutron capture radiotherapy, (ii) image and probe the internal structure of one-of-a-kind cultural heritage artifacts whilst minimizing radiation exposure inducing post-measurement material activation, and (iii) enable new methods for assessing the reliability and robustness of advanced materials and electronic components relevant to space and defense industries.

Neutron interaction with matter is weak and classic optics (e.g., mirrors \cite{KleinWerner1983} and lenses \cite{NeutronCRL, NeutronCRL2}) are ineffective for spatiotemporal modulation of illumination. Neutron beam shaping is therefore difficult and no neutron-optics equivalent of a visible-light data projector \cite{DataProjectorReference} or spatial light modulator (SLM) \cite{SLM-reference} exists. Currently a dedicated stencil (or mask) must be produced per beam shape required \cite{Grunauer2005}. A similar problem exists for x rays, however we note that a major recent advance was reported in the experimental achievement of an x-ray SLM \cite{Tamasaku2024}.

 In a conceptually quite different way, arbitrary beam shaping can be achieved using transverse displacements of a single universal patterned mask. This concept was introduced by Paganin \cite{paganin2019writing} and further developed by Ceddia {\em et al.}~\cite{ceddia2022ghostI, ceddia2022ghostII}. The method became known as {\it ghost projection} (GP), and is the reverse process of classical {\it ghost imaging} (GI) \cite{shapiro2008computational, erkmen2010ghost,Shapiro2012, Padgett2017}. Ceddia {\em et al.}~achieved the first experimental demonstration of ghost projection with x-ray synchrotron radiation in 2023 \cite{ceddia2023universal}, using a set of random masks with patterns designed in Ref.~\cite{kingston2023optimizing} and fabricated as described in Ref.~\cite{aminzadeh2023mask}.

Patterned-illumination computational neutron imaging techniques such as ghost imaging \cite{kingston2020neutron} and single-pixel camera imaging \cite{he2021single} have been successfully achieved. These realizations both provide the hardware (patterned mask and manipulation stages) and suggest ghost projection is applicable to neutrons. Here we demonstrate the ability to {\em arbitrarily shape a neutron beam using just a single (or universal) randomly patterned mask}. To perform GP, we image a mask sub-region through thousands of transverse mask displacements, then given this set of {\it basis} images and a specified target image, we determine the set of mask positions and associated exposure times that combine to produce the target image.

The paper proceeds as follows: We first describe the theory of ghost projection in Sec.~\ref{sec:background}. We then explore the capabilities and limitations of ghost projection by simulation in Sec. \ref{sec:simulations}. Given this theoretical foundation, we move to the experimental set up to achieve arbitrary beam shaping by GP at the Dingo neutron imaging beamline in Sec.~\ref{sec:method}. Images of the resulting projected neutron beams are presented in Sec.~\ref{sec:results}. This is followed by a discussion in Sec.~\ref{sec:discussion} and some concluding remarks and future research directions are presented in Sec.~\ref{sec:conclusion}.

\section{Ghost projection theory}
\label{sec:background}

\begin{figure}
    \centering
    \includegraphics[width=\linewidth]{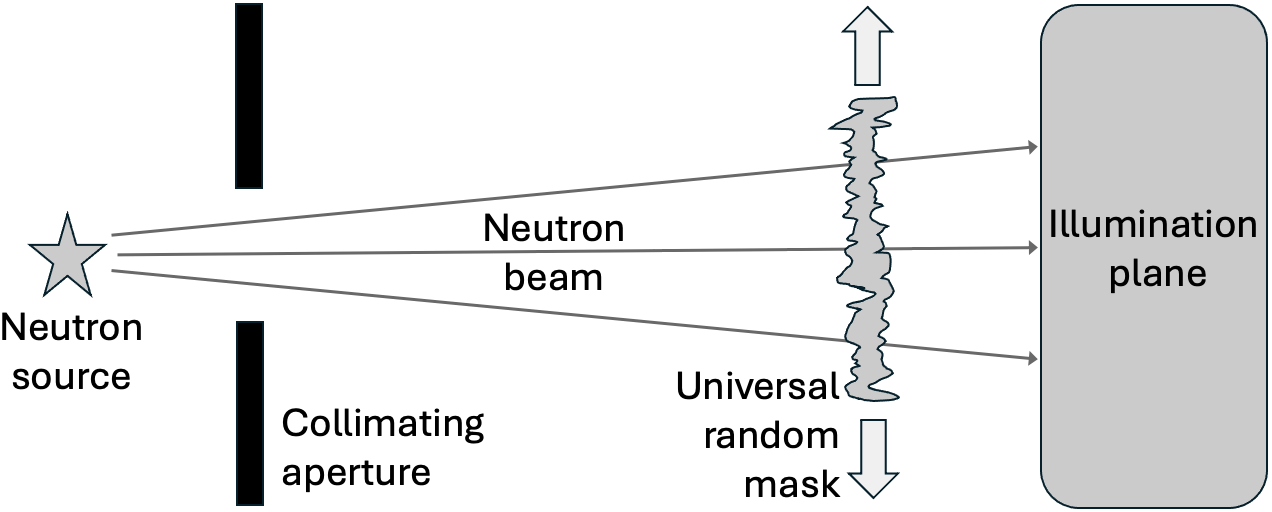}
    \caption{Schematic for neutron beam shaping by ghost projection.  By moving the universal random mask to a number of specified transverse locations, any time-integrated spatial distribution of exposure can be created over the illumination plane (up to a spatial resolution dictated by the minimum feature size created by the mask, and an additive ``pedestal'').}
    \label{fig:gpSDetup}
\end{figure}

The ghost-projection concept \cite{paganin2019writing, ceddia2022ghostI, ceddia2022ghostII} may be applied to neutrons via the schematic in Fig.~\ref{fig:gpSDetup}.  Here we see an emitter, e.g.~a nuclear-reactor \cite{KleinWerner1983} or spallation \cite{SpallationSources} source, releasing a stream of neutrons that is collimated into a beam which then passes through a single spatially-random mask. The transverse location of this mask may be precisely and reproducibly scanned in two directions, $(x,y)$, that are both transverse to the beam axis, $z$. Significant flexibility in the choice of random patterns is possible, with the key requirement being that the mask should create a set of $K$ different random intensity maps, $\{M_k(x,y)\}$, over the illumination plane, where $k=1,2,3,\cdots,K$; these intensity maps should be sufficiently finely detailed, such that the spatial resolution, $\ell$, of the target ghost-projection image, $I$, is no smaller than the finest level of detail that is present in $\{M_k(x,y)\}$ to a non-negligible degree.  For a target image with finite area $\mathcal{A}$, there will be approximately $\mathcal{N}\equiv\mathcal{A}/\ell^2$ independent resolution elements (``pixels'').  If $K$ is taken to be sufficiently large relative to $\mathcal{N}$, then the set $\{M_k(x,y)\}$ is an overcomplete \cite{MandelWolf} nonorthogonal basis in the sense that the target image may be expressed as a linear combination of random patterns drawn from $\{M_k(x,y)\}$, with nonnegative weights proportional to the exposure time for each pattern. The finite-resolution target image, $I$, will be synthesized up to an additive background termed a ``pedestal''. 
Crucially, the overcomplete nature of the random-pattern basis implies that {\em most of the assigned weights can be set to zero}; the larger $K$ is chosen to be (relative to $\mathcal{N}$), the larger the fraction of the basis that may be discarded for the purposes of creating a single specified target pattern of time-integrated exposure. In this scenario, the ghost-projection process becomes more efficient since the total number of nonzero weights is a nonincreasing function of increasing $K$.  A different subset of the random-pattern basis may, and in general should, be chosen for each target ghost-projection image.
 
\begin{figure}
    \centering
    \includegraphics[width=\linewidth]{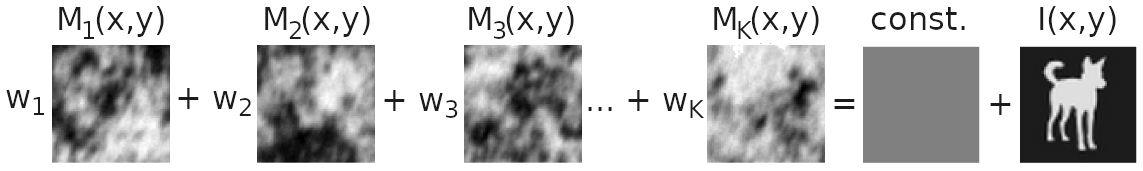}
    \caption{A depiction of the ghost projection principle, constructing a desired image (plus a pedestal) from the non-negative weighted sum of $K$ patterned illuminations.}
    \label{fig:NGP_process_diagram}
\end{figure}

Given our set of $K$ basis illumination patterns provided by the mask subject to various transverse displacements, these patterns are vectorized and collated into a matrix, $M$. For ghost projection we require a vector of $K$ non-negative weights, $w$, that produce a vectorized target image, $I$, as follows:
\begin{equation}\label{eqn:GpProblem}
    Mw = I.
\end{equation}
Here, as mentioned earlier, the weights represent illumination exposure times and nonnegative exposures are required to be physically realizable. A practical depiction of this mathematical description is given in Fig.~\ref{fig:NGP_process_diagram}. Several approaches have been proposed to determine the weights, $w$, as correlation values, correlation filtration, through nonnegative least squares optimization, or L1-norm minimization with nonnegative regularization (see e.g., \cite{paganin2019writing, ceddia2022ghostI}). 
The weights can be determined by solving the non-negative least-squares problem:
\begin{equation}\label{eqn:GpOptimisationProb}
   \text{arg~min} \|Mw - I\|,\text{~subject to~} w_k \geq 0~\forall~k \in [1,K].
\end{equation}
The nonnegative weights, $w$, are rescaled to per-mask exposure times according to the application required. For additional mathematical development, see Refs.~\cite{paganin2019writing, ceddia2022ghostI, ceddia2022ghostII}.

There are two distinct contexts where ghost projection may be useful, which lead to differences in the specific formulation of the least-squares problem \eqref{eqn:GpOptimisationProb}.
Firstly, if dose is a significant consideration, then the columns of $M$ should be flattened copies of each sub-mask, and the vector $I$ should be a flattened copy of the target image.
Secondly, if dose is not a significant consideration but rather the fidelity of the projected image, then the mask matrix $M$ and target vector $I$ can be mean-corrected. This allows the projection to capture more features in the target image, at the expense of introducing a constant offset dose (or pedestal). In this case, the weights are calculated by
\begin{align}
    \text{arg~min} \|\widehat{M}w - \widehat{I}\|,\text{~subject to~} w_k \geq 0~\forall~k \in [1,K], \label{eqn:GpOptimisationProbMeanAdj}
\end{align}
where $\widehat{M}=M-\overline{M}$ and $\widehat{I}=I-\overline{I}$, for mean values $\overline{M}$ and $\overline{I}$.

\begin{figure}
    \centering
    \begin{minipage}{0.33\linewidth}
        \centering
        \scriptsize{(a)}\\
        \includegraphics[width=0.95\linewidth]{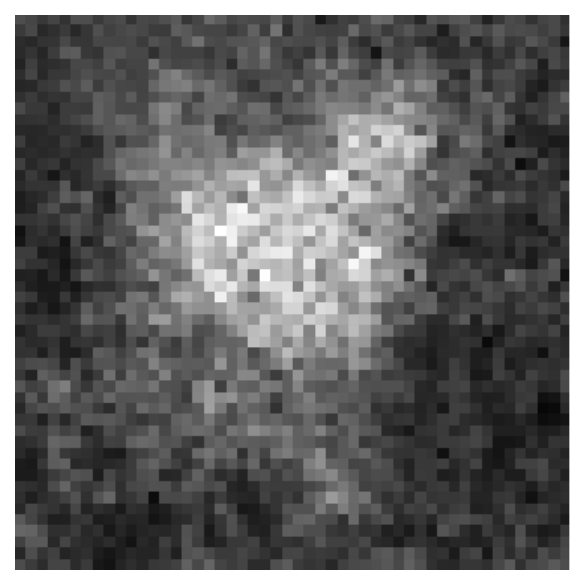}
    \end{minipage}%
    \begin{minipage}{0.33\linewidth}
        \centering
        \scriptsize{(b)}\\
        \includegraphics[width=0.95\linewidth]{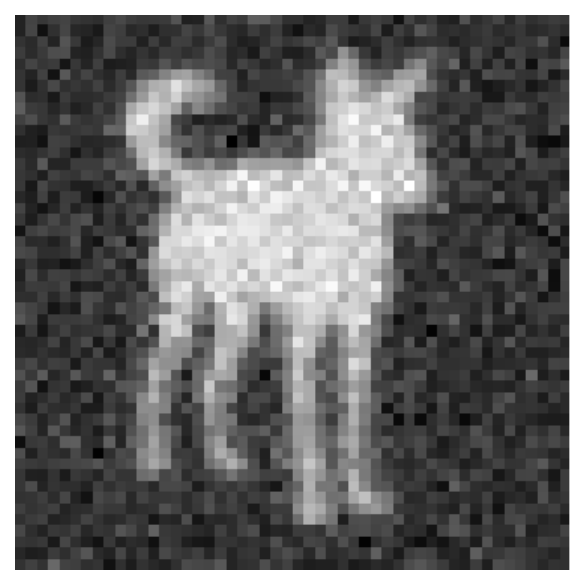}
    \end{minipage}%
    \begin{minipage}{0.33\linewidth}
        \centering
        \scriptsize{(c)}\\
        \includegraphics[width=0.95\linewidth]{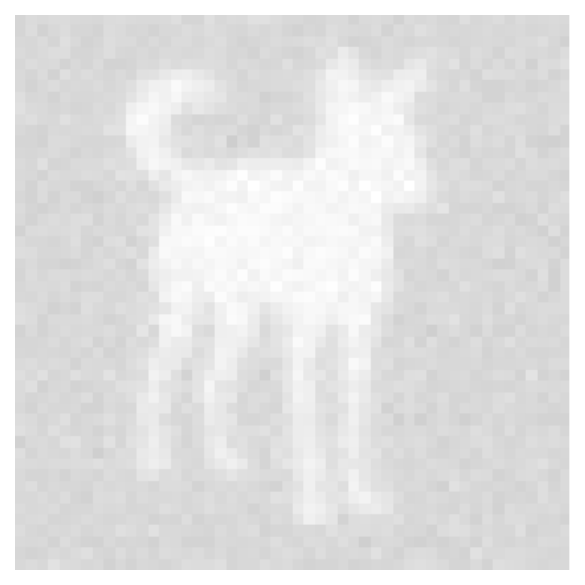}
    \end{minipage}\\
    \caption{Computational projections of $50\times 50$ pixel version of Fig.~\ref{fig:targetImages}(f) using large fractal mask Fig.~\ref{fig:GdMask}(b) scanned over a $50\times 50$ grid of positions: (a) no mean adjustment, true scale; (b) with mean adjustment, normalized scale; (c) with mean adjustment, true scale.}
    \label{fig:reconsDemo}
\end{figure}

These two options are illustrated by simulation in Fig.~\ref{fig:reconsDemo}. Here, a $50\times 50$ pixel target image is reconstructed using 2500 patterns (detailed in the following simulation section). All weights were determined using the non-negative least-squares solver BLLS from the GALAHAD \cite{gould2003galahad} optimization library. Applying no mean adjustment (i.e.~the dose-sensitive case) by solving \eqref{eqn:GpOptimisationProb} gives the projected image Fig.~\ref{fig:reconsDemo}(a), while applying mean adjustment gives the projected image shown in Fig.~\ref{fig:reconsDemo}(b) and (c); the image in (b) is normalized so as to not show the pedestal, while (c) shows the true projected image with the pedestal.

In the next section we explore, by simulation, the effect of pedestal requirements on the ghost projection image fidelity and contrast attainable. The importance of the pedestal also dictates the number and type of patterns selected for ghost projection.

\section{Ghost projection simulations}
\label{sec:simulations}

\begin{figure}
    \centering
    \begin{minipage}{0.25\linewidth}
        \centering
        \scriptsize{(a)}\\
        \includegraphics[width=0.95\linewidth]{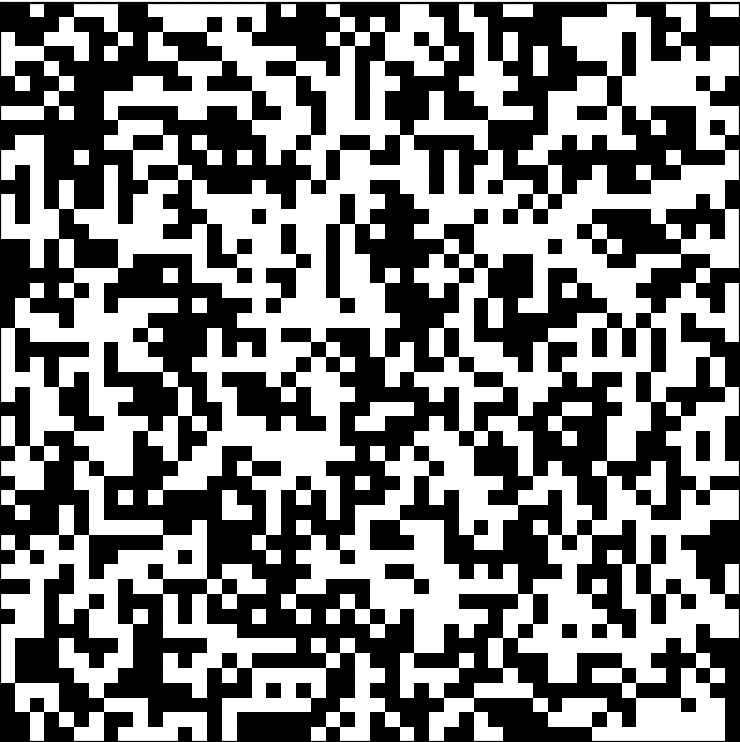}
    \end{minipage}%
    \begin{minipage}{0.25\linewidth}
        \centering
        \scriptsize{(b)}\\
        \includegraphics[width=0.95\linewidth]{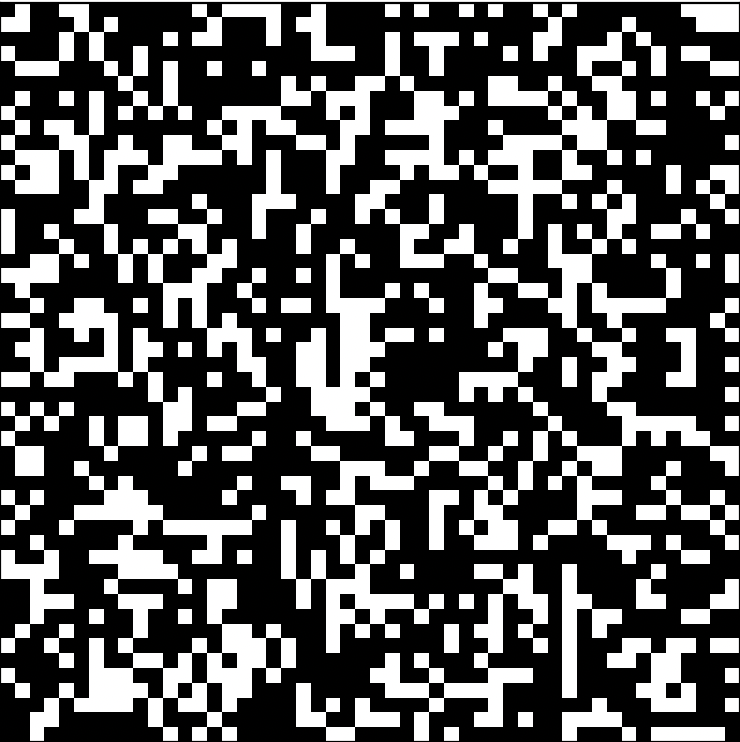}
    \end{minipage}%
    \begin{minipage}{0.25\linewidth}
        \centering
        \scriptsize{(c)}\\
        \includegraphics[width=0.95\linewidth]{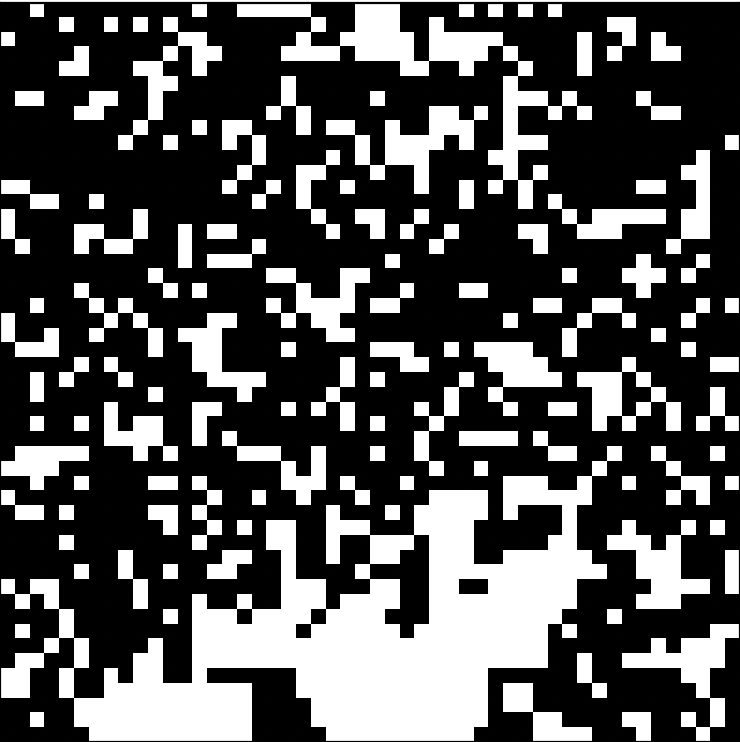}
    \end{minipage}%
    \begin{minipage}{0.25\linewidth}
        \centering
        \scriptsize{(d)}\\
        \includegraphics[width=0.95\linewidth]{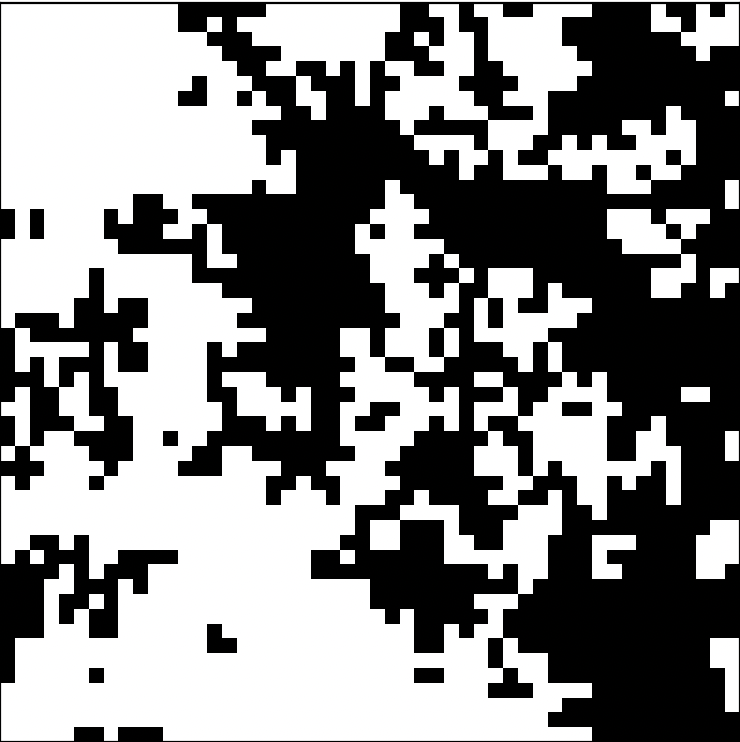}
    \end{minipage}\\
    \caption{Example $50 \times 50$ pixel, binary patterns used for the simulations. (a) random, (b) small fractal, (c) medium fractal, (d) large fractal. (see Sec. \ref{sec:method} for more detail).}
    \label{fig:simMasks}
\end{figure}

In order to appreciate the simulation results, we must first describe the set of patterns considered. We have employed the patterns from the designed scanning mask used in the neutron experiments outlined below. There are four types of masks with different degrees of fractality (see Sec. \ref{sec:method} for more detail). Fractal patterns have been shown to be beneficial in novel patterned illumination experiments when the precise field-of-view and resolution attainable are not known beforehand \cite{kingston2023optimizing}. Each mask is 10 mm $\times$ 10 mm with 10 $\mu$m pixel size. We have simulated a 0.5 mm $\times$ 0.5 mm scanning window (i.e., $50 \times 50$ pixels) with a scanning stride of 19 pixels both horizontally and vertically. This gives a $50 \times 50$ grid of patterns per mask type, the same as in the experiment. An example of each pattern is presented in Fig. \ref{fig:simMasks}. Note that this is not an overcomplete set for each pattern type as is desirable for ghost projection, but the simulations will demonstrate the principles and what is possible experimentally given a similar set.

These sets of patterns are flattened to form the columns of $M$ and we have used the silhouette of a dingo as our target image, $I$. As noted in Sec.~\ref{sec:background}, the weights, $w$, determined by solving \eqref{eqn:GpOptimisationProb} have low total dose, but (excluding the pedestal) less fidelity to the target image than weights determined by solving the mean-adjusted problem \eqref{eqn:GpOptimisationProbMeanAdj}. This trade-off motivates a continuum of non-negative least-squares problems to determine the weights, namely
\begin{align}
    w(\alpha) = \text{arg~min} \|(M-\alpha\overline{M})w - (I-\alpha\overline{I})\|,\text{~s.t.~} w_k \geq 0,  \label{eqn:GpOptimisationProbContinuum}
\end{align}
where $\alpha\in[0,1]$ is a parameter, with $\alpha=0$ and $\alpha=1$ yielding \eqref{eqn:GpOptimisationProb} and \eqref{eqn:GpOptimisationProbMeanAdj} respectively.

\begin{figure}
    \centering
    \begin{minipage}{0.33\linewidth}
        \centering
        \scriptsize{(a)}\\
        \includegraphics[width=0.95\linewidth]{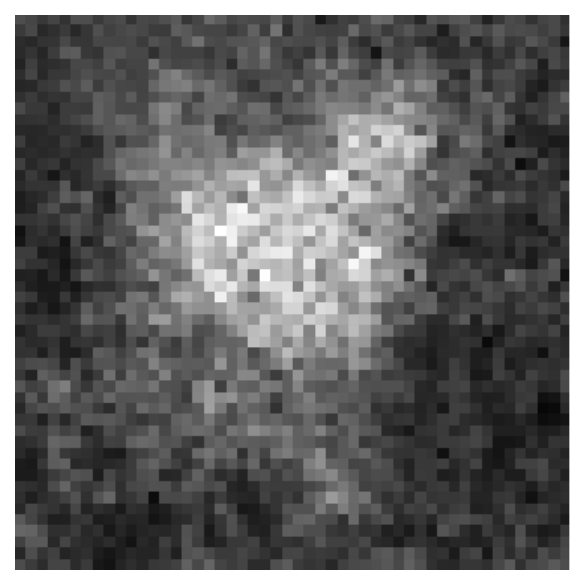}
    \end{minipage}%
    \begin{minipage}{0.33\linewidth}
        \centering
        \scriptsize{(b)}\\
        \includegraphics[width=0.95\linewidth]{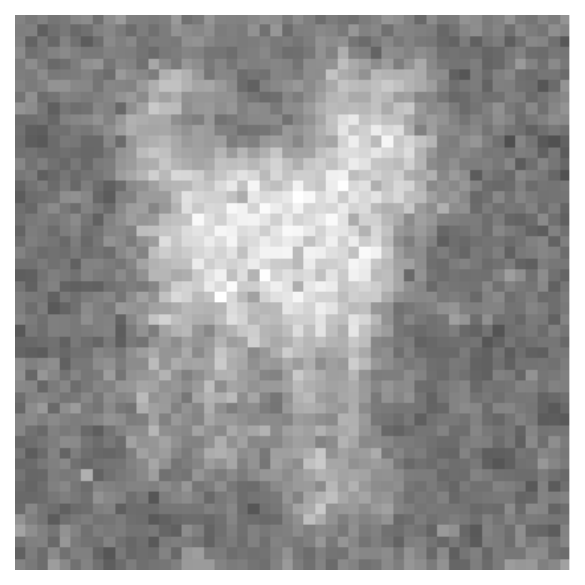}
    \end{minipage}%
    \begin{minipage}{0.33\linewidth}
        \centering
        \scriptsize{(c)}\\
        \includegraphics[width=0.95\linewidth]{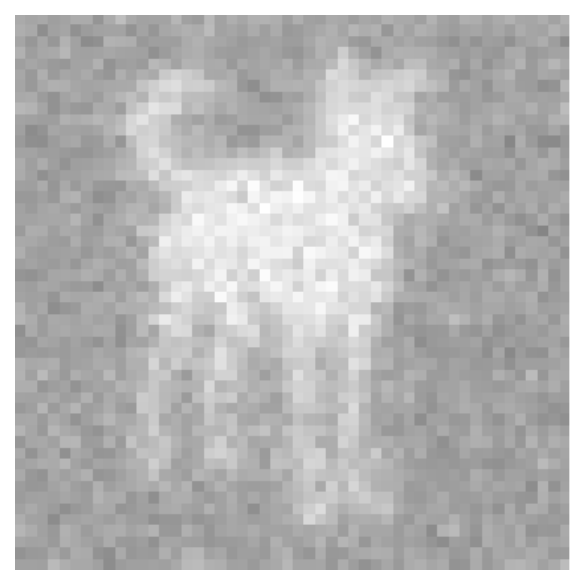}
    \end{minipage}\\
    \begin{minipage}{0.33\linewidth}
        \centering
        \scriptsize{(d)}\\
        \includegraphics[width=0.95\linewidth]{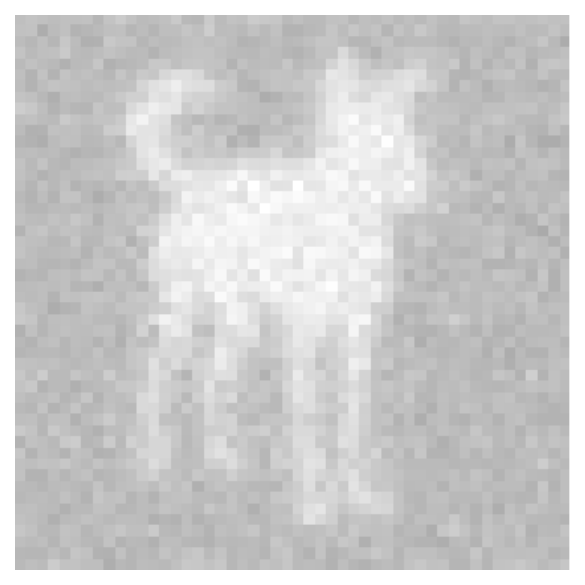}
    \end{minipage}%
    \begin{minipage}{0.33\linewidth}
        \centering
        \scriptsize{(e)}\\
        \includegraphics[width=0.95\linewidth]{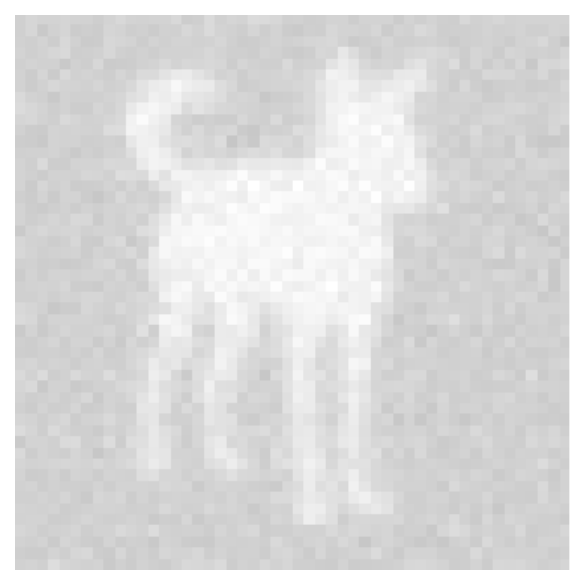}
    \end{minipage}%
    \begin{minipage}{0.33\linewidth}
        \centering
        \scriptsize{(f)}\\
        \includegraphics[width=0.95\linewidth]{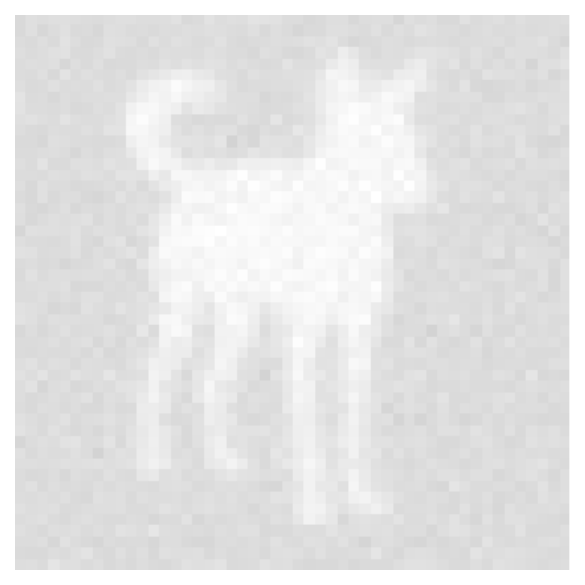}
    \end{minipage}\\
    \caption{Computational projections of $50\times 50$ pixel version of Fig.~\ref{fig:targetImages}(f) using large fractal mask Fig.~\ref{fig:GdMask}(b) scanned over a $50\times 50$ grid of positions. Weights come from solving \eqref{eqn:GpOptimisationProbContinuum} with mean adjustment (a) $\alpha=0$. (b) $\alpha=0.2$. (c) $\alpha=0.4$. (d) $\alpha=0.6$. (e) $\alpha=0.8$. (f) $\alpha=1$.}
    \label{fig:pedestalVary}
\end{figure}

Figure~\ref{fig:pedestalVary} shows the projections achieved for the same problem as Fig.~\ref{fig:reconsDemo}, but with the mean adjustment level $\alpha$ in \eqref{eqn:GpOptimisationProbContinuum} varying between 0 and 1. The final weights are scaled by a constant factor to ensure the projected image has unit maximum intensity. Visually, it is clear that larger $\alpha$ improves the resolution and fidelity (ignoring the pedestal), while increasing the total dose.

\begin{figure}
    \centering
    \begin{minipage}{0.49\linewidth}
        \centering
        \scriptsize{(a)}\\
        \includegraphics[width=0.95\linewidth]{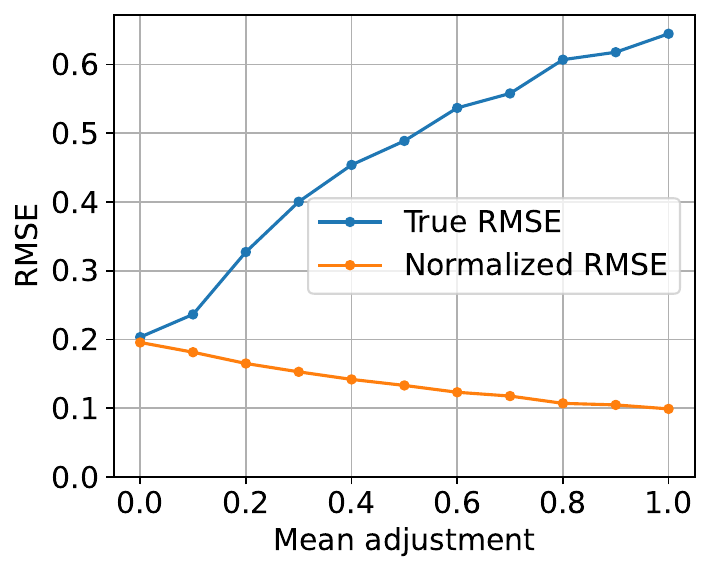}
    \end{minipage}%
    \begin{minipage}{0.49\linewidth}
        \centering
        \scriptsize{(b)}\\
        \includegraphics[width=0.95\linewidth]{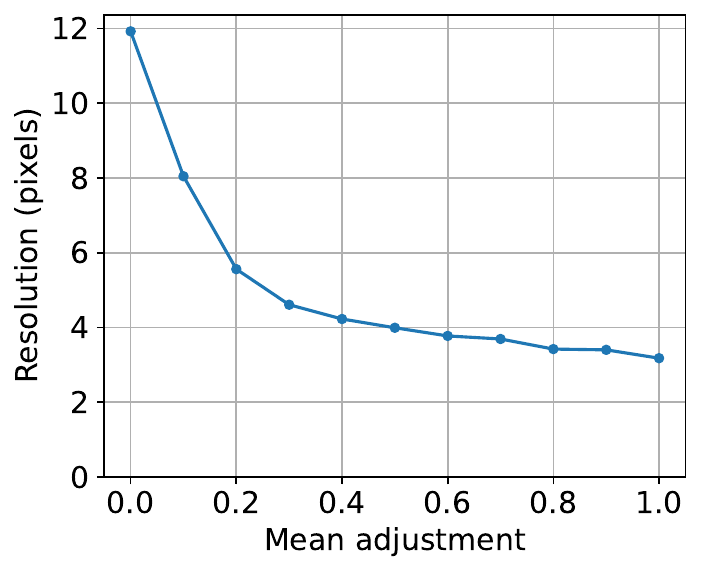}
    \end{minipage}\\
    \begin{minipage}{0.49\linewidth}
        \centering
        \scriptsize{(c)}\\
        \includegraphics[width=0.95\linewidth]{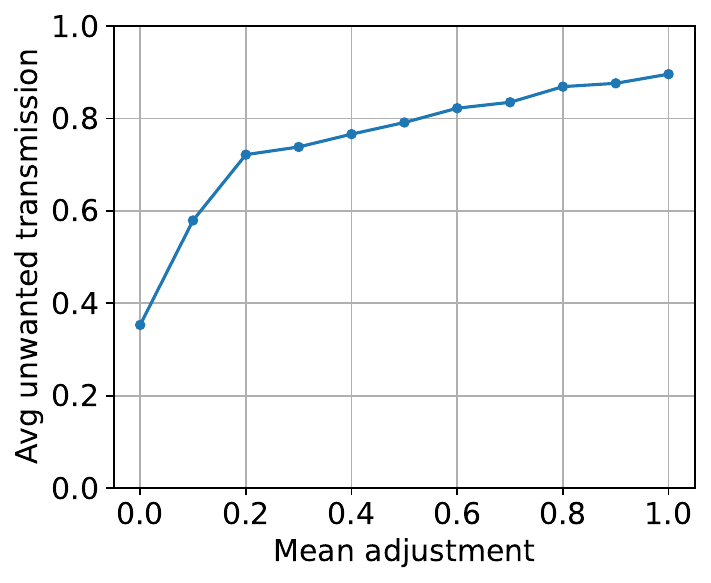}
    \end{minipage}%
    \begin{minipage}{0.49\linewidth}
        \centering
        \scriptsize{(d)}\\
        \includegraphics[width=0.95\linewidth]{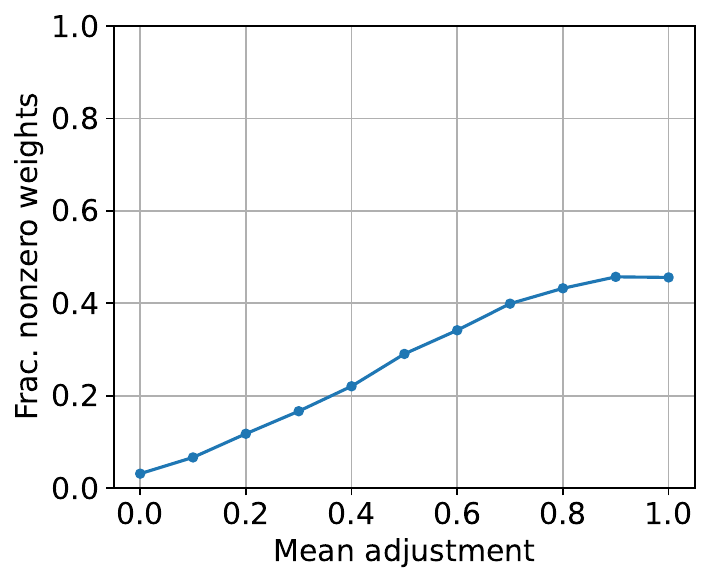}
    \end{minipage}\\
    \caption{Impact of varying mean adjustment size for computational projections of $50\times 50$ pixel version of Fig.~\ref{fig:targetImages}(f) using large fractal mask Fig.~\ref{fig:GdMask}(b) scanned over a $50\times 50$ grid of positions. (a) RMSE. (b) Resolution. (c) Average unwanted transmission. (d) Fraction nonzero weights.}
    \label{fig:pedestalVaryMetrics}
\end{figure}

To assess the impact of varying $\alpha$, the following metrics were used to compare the projected image with the target image:
\begin{itemize}
    \item {\bf RMSE} (Root mean square error) both ``true'' (i.e.~including the impact of the pedestal) and ``normalized'' (i.e.~ignoring the pedestal), where the target and projected images have the mean subtracted and the projection image is scaled to minimize RMSE;
    \item {\bf Resolution} determined using Fourier ring correlation between the target and projected images \cite{saxton1982correlation, vanHeel1982arthropod}. Here we have employed the conservative 1-bit threshold defined in \cite{van2005fourier};
    \item {\bf Average ``unwanted'' transmission} defined as the average transmitted intensity of the projected image in areas where the target image specifies no transmission. This in effect measures the total dose in regions where no dose is expected;
    \item {\bf Fraction of nonzero weights}, $w$; ``nonzero'' defined here as $w_k > 10^{-3}$.
\end{itemize}
The values of these metrics applied to the projections in Fig.~\ref{fig:pedestalVary} are shown in Fig.~\ref{fig:pedestalVaryMetrics}. As expected, larger values of $\alpha$ (i.e.~larger pedestals) improve the resolution and normalized RMSE, while increasing the true RMSE (which includes the pedestal), amount of unwanted transmission, and the fraction of nonzero weights.

\begin{figure}
    \centering
    \begin{minipage}{0.33\linewidth}
        \centering
        \scriptsize{(a)}\\
        \includegraphics[width=0.95\linewidth]{./figures/simulations/recons_largeFrac_dingoSil_no_pedestal.pdf}
    \end{minipage}%
    \begin{minipage}{0.33\linewidth}
        \centering
        \scriptsize{(b)}\\
        \includegraphics[width=0.95\linewidth]{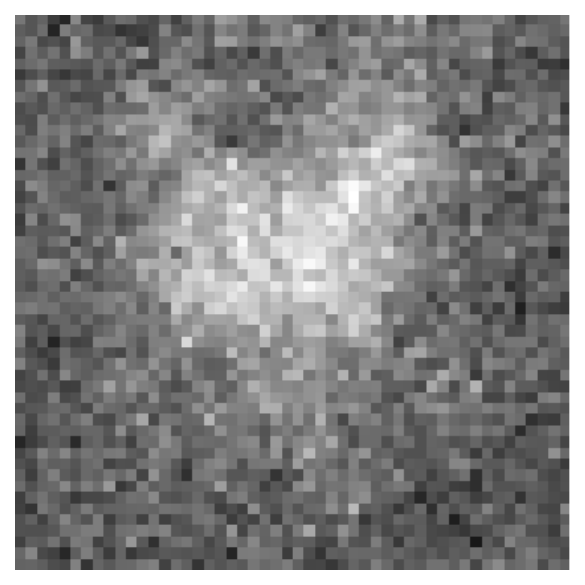}
    \end{minipage}%
    \begin{minipage}{0.33\linewidth}
        \centering
        \scriptsize{(c)}\\
        \includegraphics[width=0.95\linewidth]{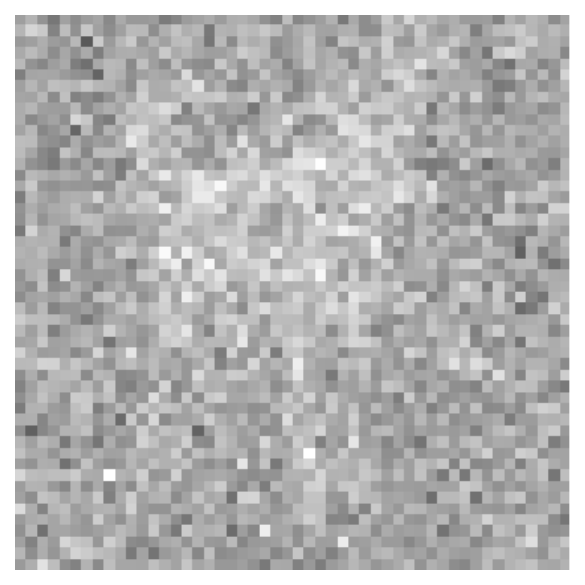}
    \end{minipage}\\
    \begin{minipage}{0.33\linewidth}
        \centering
        \scriptsize{(d)}\\
        \includegraphics[width=0.95\linewidth]{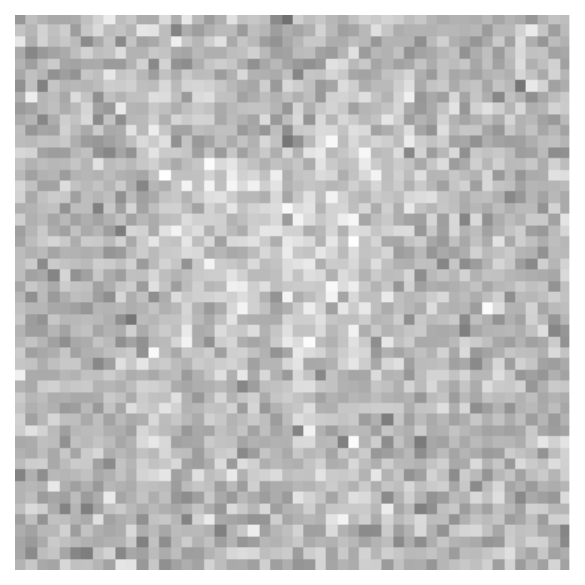}
    \end{minipage}%
    \begin{minipage}{0.33\linewidth}
        \centering
        \scriptsize{(e)}\\
        \includegraphics[width=0.95\linewidth]{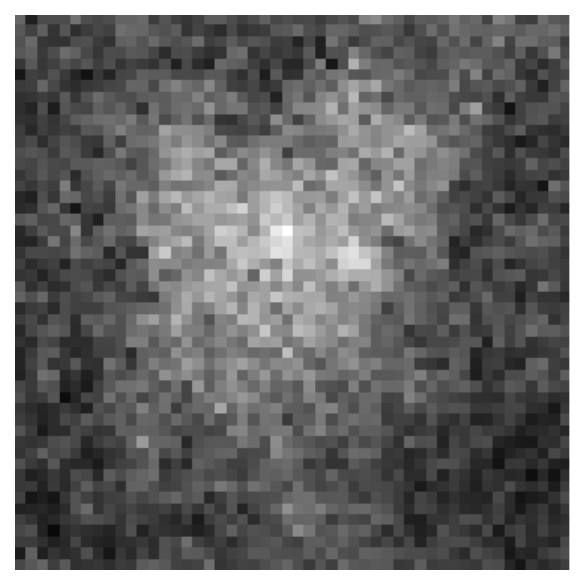}
    \end{minipage}%
    \begin{minipage}{0.33\linewidth}
        \centering
        \scriptsize{(f)}\\
        \includegraphics[width=0.95\linewidth]{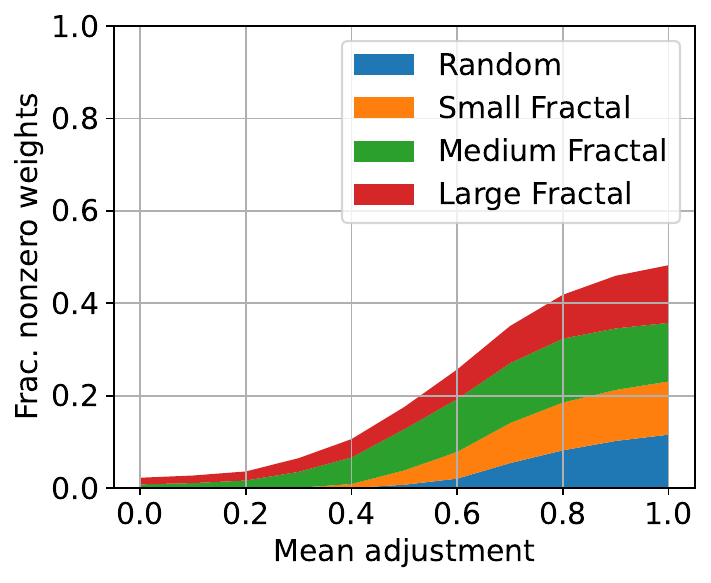}
    \end{minipage}\\
    \caption{Computational projections of $50\times 50$ pixel version of Fig.~\ref{fig:targetImages}(f) without mean adjustment, with mask scanned over a $50\times 50$ grid of positions. (a) Large fractal mask. (b) Medium fractal mask. (c) Small fractal mask. (d) Random mask. (e) Mixed mask (25\% of each type). (f) Fraction of mask types selected as mean adjustment varies.}
    \label{fig:reconsMaskVary}
\end{figure}

Up to this point, the simulations all used the set of patterns from the large fractal mask only. The choice of mask type also affects the quality of the projection. Figures~\ref{fig:reconsMaskVary}(a)--(d) show the same projection as Fig.~\ref{fig:reconsDemo}(a), with weights determined via \eqref{eqn:GpOptimisationProb}, i.e.~$\alpha=0$ in \eqref{eqn:GpOptimisationProbContinuum}, for masks with (a) large, (b) medium, and (c) small fractal structures, and (d) purely random masks. Using masks with larger fractal structures better captures the low frequency structures in the target, but smaller fractal structures (or random masks) better capture the high frequency structures in the image (such as the dingo's legs). A ``mixed''-type mask based on a random subset of 25\% of all masks of each type gave the projection in Fig.~\ref{fig:reconsMaskVary}(e), which captures more of the high frequency features than using purely large fractal masks, with a trade-off of worse reconstruction of the low frequency features.

In the case of a ``mixed'' mask, the size of $\alpha$ in \eqref{eqn:GpOptimisationProbContinuum} affects which types of masks are selected. In Fig.~\ref{fig:reconsMaskVary}(f), the fraction of nonzero masks as $\alpha$ varies from 0 to 1 is shown, colored by mask type. For low values of $\alpha$, when a high true-fidelity but low-resolution projection is desired, large and medium fractal masks are selected. For large values of $\alpha$, where a high-resolution and normalized-fidelity projection is desired, all mask types are selected.

Having established the ghost projection concepts and explored the expected capabilities in image representation by simulation, we will now turn to experimental demonstration of beam shaping with neutrons by ghost projection.

\section{Experimental Method}
\label{sec:method}

The experiments were performed using the open-pool reactor-based neutron source on the Dingo neutron imaging beamline at the Australian Centre for Neutron Scattering (ACNS) \cite{Garbe2015,Garbe2017}.  A polyenergetic neutron beam was employed. Epithermal neutrons were removed by a 90 mm$\,\times\,$90 mm$\,\times\,$30 mm sapphire filter composed of several superoptical quality crystals with the [001] axis parallel to the incoming beam; the resulting thermal neutron spectrum had a maximum intensity at wavelength $1.5$\AA. The detector was positioned $L=9.8$m from a diameter $d=19.8$ mm pin-hole at the neutron source, giving a beam divergence \cite{Treimer2009} of $\Theta = d/L= 1/495$. A two-section 4.5 m long flight tube filled with He at ambient pressure (1 bar) was used to reduce neutron scatter from air. In this high-intensity configuration, the average neutron radiation flux was $4.7 \times 10^7 \textrm{n}. \textrm{cm}^{-2} \textrm{s}^{-1}$ \cite{jakubowski2023monte}.

\begin{figure}
    \centering
    \begin{minipage}{0.5\linewidth}
        \centering
        \scriptsize{(a)}\\
        \includegraphics[width=0.9\linewidth]{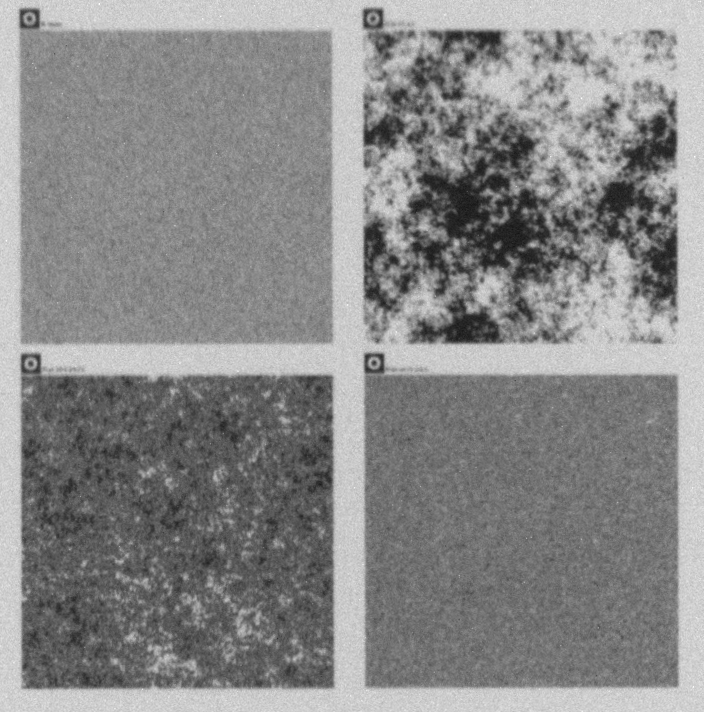}
    \end{minipage}%
    \begin{minipage}{0.5\linewidth}
        \centering
        \scriptsize{(b)}\\
        \includegraphics[width=0.9\linewidth]{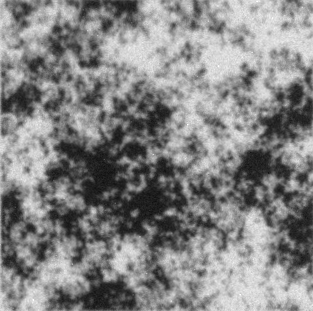}
    \end{minipage}
    \caption{(a) Normalized neutron radiograph of the Gd mask designed and fabricated for ghost imaging. Each fractal mask has dimensions of 10 mm $\times$ 10 mm. (b) The large fractal-mask used to create illumination patterns for ghost projection.}
    \label{fig:GdMask}
\end{figure}

Figure \ref{fig:GdMask}(a) depicts the four masks that we designed and fabricated for ghost imaging and ghost projection experiments. These images were acquired as 10 accumulations each with 12s exposure time and a 30$\mu$m thick gadolinium oxysulfide (Gd$_2$O$_2$S) scintillation screen imaged with a 32$\mu$m pixel size. The top-left mask in Fig.~\ref{fig:GdMask}(a) is a random binary mask with uncorrelated pixel values, while the other three are statistically-fractal random binary masks with three different degrees of fractality. We refer to these as small, medium, and large fractal masks based on their feature sizes that are determined by the critical exponent governing the power-law decay, $\alpha = -0.5, -0.75$ and $-1$, with regularization $\beta = 0.25~\mu m^{-1}$ for all masks. The mathematical description of these fractal masks can be found in Refs.~\cite{kingston2023optimizing, ceddia2023universal}. Only the large fractal mask, shown in Fig.~\ref{fig:GdMask}(b), was used for our neutron ghost projection experiments. The size and the resolution of each mask are 10 mm $\times$ 10 mm and 10 $\mu$m respectively. The fabrication process was performed at Paul Scherrer Institute (PSI) by lasering the patterns into a 5 $\mu$m Gd layer, which was coated on a glass substrate. 

A mask manipulator stage was designed and constructed at the Australian Nuclear Science and Technology Organisation (ANSTO). It consists of two high-precision translation stages to manipulate the mask transverse to the neutron beam direction ({\it xfine} and {\it yfine}) and a coarse stage, {\it xghost}, to bring the whole assembly into and out of the neutron beam. {\it xghost} is a simple belt drive with a rotary absolute encoder employed on the belt drive shaft. {\it xfine} and {\it yfine} have a $\pm$30 mm translation range, driven by a lead screw and stepper motor with 40 nm steps. They are encoded by linear absolute encoders supplied by Fagor. The encoders have a range of up to 70 mm, are compatible with the standard Galil controllers used at ANSTO, and give the transverse mask position in the $(x,y)$-plane with nanometer precision.

\begin{figure}
    \centering
    \includegraphics[width=0.9\linewidth]{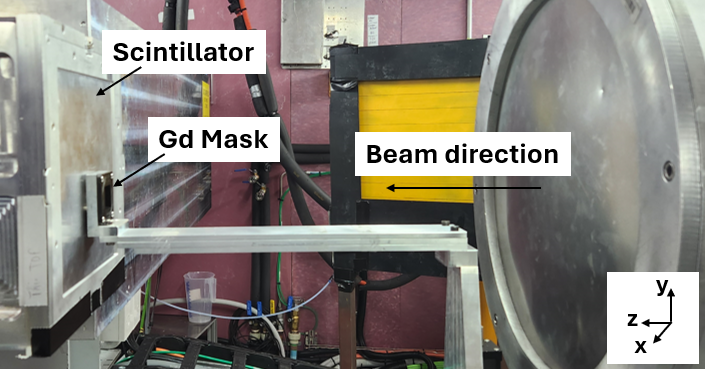}
    \caption{Experimental setup for arbitrary-profile neutron beam shaping via a universal ghost-projection mask.}
    \label{fig:expSetup}
\end{figure}

A photograph of the experimental setup is shown in Fig.~\ref{fig:expSetup}. The Gd mask was mounted on an aluminum arm and positioned as close as possible to the scintillator to minimize pattern degradation from beam divergence. The mask was moved in the $x$ and $y$ directions (perpendicular to the beam direction $z$), and the radiographs of the mask at different transverse locations were collected at the scintillator. 

The detector consisted of a $^6$LiF/ZnS:Cu scintillation screen of thickness 50 $\mu$m, a mirror, and a ZWO ASI2600MM PRO Cooled camera, with a back-illuminated 16-bit CMOS sensor, that was placed out of the neutron beam. The CMOS camera has a $4176 \times 6248$ pixel array that was coupled with a Carl Zeiss 100 mm fixed focal length lens to achieve an effective pixel size of 15.6 $\mu$m.

High-intensity neutron flux with greater divergence (i.e., lower resolution) was selected as the experimental conditions to ensure signal-to-noise ratio was not an issue. Under these conditions, only the pattern with the largest features, $\alpha = -1$, had any resolvable features. We were therefore limited to using only this mask for the present ghost projection demonstration. The mask was scanned with a 2 mm $\times$ 2 mm field of view over a $50 \times 50$ grid of positions. The spacing between rows and columns of positions was 160 $\mu$m to span the entire 10 mm $\times$ 10 mm large fractal mask (see Fig.~\ref{fig:GdMask}(b)). At each position a 15s exposure radiograph was recorded in high-intensity mode to increase signal-to-noise ratio. Five open-beam and dark images were also recorded to normalize the basis set of images and remove the effects of dark current, beam profile, and scintillator structure variations.

\begin{figure}
    \centering
    \begin{minipage}{0.33\linewidth}
        \centering
        \scriptsize{(a)}\\
        \includegraphics[width=0.95\linewidth]{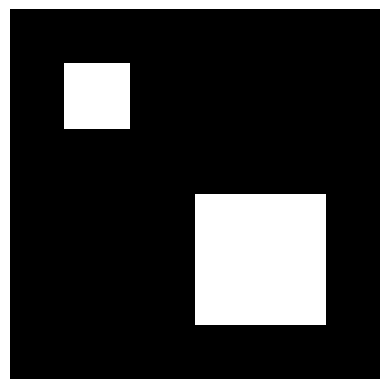}
    \end{minipage}%
    \begin{minipage}{0.33\linewidth}
        \centering
        \scriptsize{(b)}\\
        \includegraphics[width=0.95\linewidth]{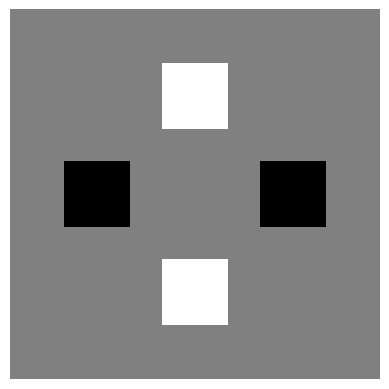}
    \end{minipage}%
    \begin{minipage}{0.33\linewidth}
        \centering
        \scriptsize{(c)}\\
        \includegraphics[width=0.95\linewidth]{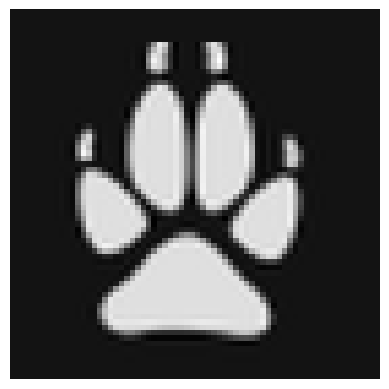}
    \end{minipage}\\
    \begin{minipage}{0.33\linewidth}
        \centering
        \scriptsize{(d)}\\
        \includegraphics[width=0.95\linewidth]{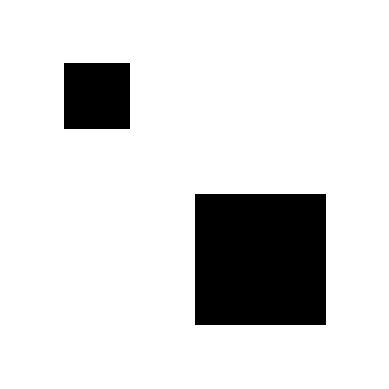}
    \end{minipage}%
    \begin{minipage}{0.33\linewidth}
        \centering
        \scriptsize{(e)}\\
        \includegraphics[width=0.95\linewidth]{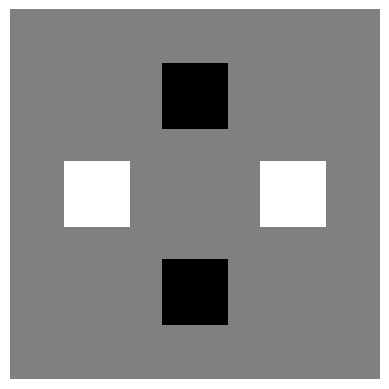}
    \end{minipage}%
    \begin{minipage}{0.33\linewidth}
        \centering
        \scriptsize{(f)}\\
        \includegraphics[width=0.95\linewidth]{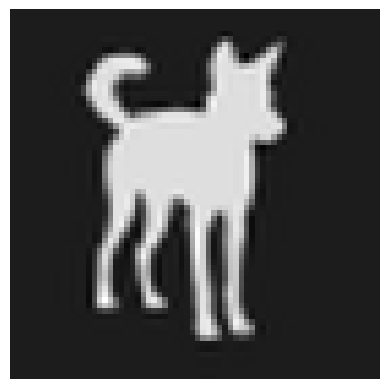}
    \end{minipage}
    \caption{The target images designed to demonstrate arbitrary-profile neutron beam shaping. Each image has physical dimensions 2 mm $\times$ 2 mm represented as $68 \times 68$ pixels.}
    \label{fig:targetImages}
\end{figure}

As shown in Fig.~\ref{fig:targetImages}, six target images were designed to demonstrate beam shaping. They included simple squares of increased or decreased intensity relative to the background, and some more complex shapes.

Given the patterned-illumination basis set and the target images, the \texttt{scipy} nonnegative least squares optimizer was employed to solve for the weights. These weights were rescaled to have a maximum of 10 and then rounded to integer values. We set a minimum exposure time (15s in this case, as used for the mask scanning) and the integer weight specifies the number of exposures to accumulate for a given mask position. A list of mask positions was produced to generate the target image; a mask position requiring $N$ exposures was simply listed $N$ consecutive times.

\section{Experimental results and analysis}
\label{sec:results}

Of the 2500 mask positions utilized, 2307 basis patterns were recorded in total, (images 211-403 were lost in data transfer). Fourier ring correlation with the ideal designed mask subsets revealed that the resolution of these pattern radiographs is approximately 140 $\mu$m (using the 1-bit threshold defined in \cite{van2005fourier}). This set of patterns was used to generate the target images. This process and some example basis images are presented in Fig.~\ref{fig:NGP_process_diagram}.



The six target images each required around 210 mask positions (the remaining mask positions had zero weight). This was reduced to around 170 positions once the weights were modified to be integers in [0,10] since mask positions with optimized weights $<$ 0.5 were truncated. Multiple exposures were achieved by repeating the mask positions listed and the resulting number of listed positions for each target image was approximately 500.

As an example, the target image in Fig.~\ref{fig:targetImages}(b) required 427 mask positions (175 unique positions). The process of constructing the projection image after accumulating every sixth of the stage positions, i.e., every 71.16 mask positions, is presented in Fig.~\ref{fig:gpSequence}. A video showing this target image construction pattern by pattern is included in the Supplementary Material \cite{suppVideoTwoUpTwoDown} along with videos showing the construction of the other five target images \cite{suppVideoTwoSquares, suppVideoTwoSquaresNeg, suppVideoTwoUpTwoDownNeg, suppVideoDingoPaw, suppVideoDingoSil}.
\begin{figure*}
    \centering
    \begin{minipage}{0.16\linewidth}
        \centering
        \scriptsize{(a) NRMSE = 0.169}\\
        \includegraphics[width=0.95\linewidth]{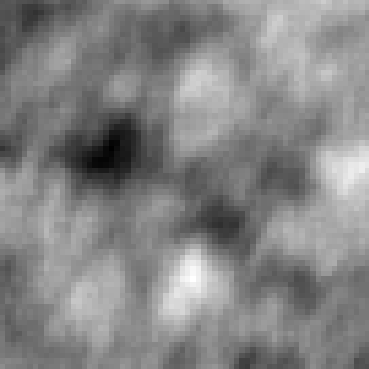}
    \end{minipage}%
    \begin{minipage}{0.16\linewidth}
        \centering
        \scriptsize{(b) NRMSE = 0.175}\\
        \includegraphics[width=0.95\linewidth]{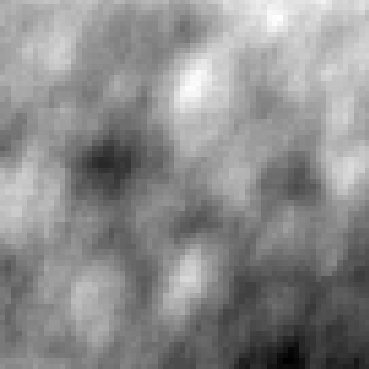}
    \end{minipage}%
    \begin{minipage}{0.16\linewidth}
        \centering
        \scriptsize{(c) NRMSE = 0.176}\\
        \includegraphics[width=0.95\linewidth]{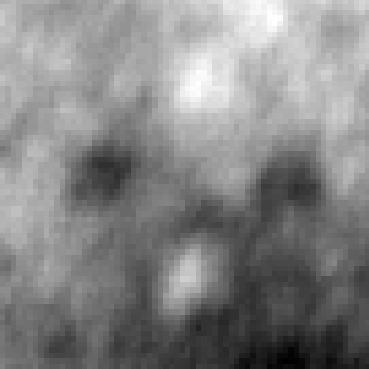}
    \end{minipage}%
    \begin{minipage}{0.16\linewidth}
        \centering
        \scriptsize{(d) NRMSE = 0.169}\\
        \includegraphics[width=0.95\linewidth]{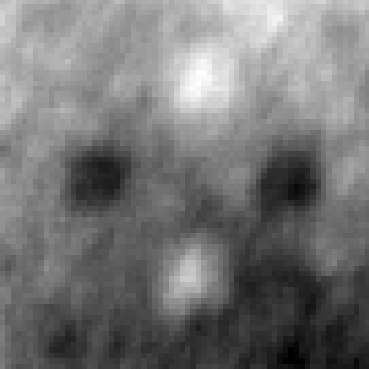}
    \end{minipage}%
    \begin{minipage}{0.16\linewidth}
        \centering
        \scriptsize{(e) NRMSE = 0.132}\\
        \includegraphics[width=0.95\linewidth]{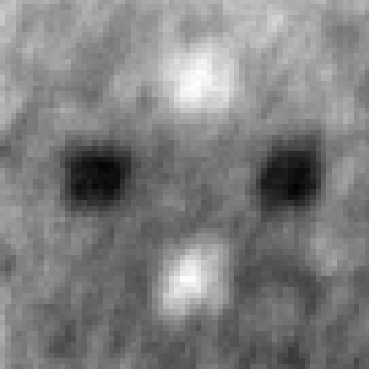}
    \end{minipage}%
    \begin{minipage}{0.16\linewidth}
        \centering
        \scriptsize{(f) NRMSE = 0.112}\\
        \includegraphics[width=0.95\linewidth]{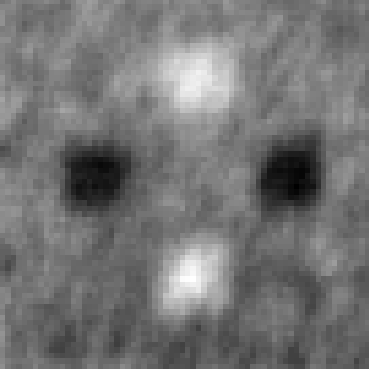}
    \end{minipage}
    \caption{The process of constructing the projection of the second target image (Fig.~\ref{fig:targetImages}(b)) after summing over (a) 71, (b) 142, (c) 213, (d) 284, (e) 355, and (f) 427 of the total 427 stage positions computed. Each image has physical dimensions 2 mm $\times$ 2 mm represented as $68 \times 68$ pixels. A corresponding video is given in the Supplementary Material \cite{suppVideoTwoUpTwoDown}. Normalized RMSE (NRMSE) is reported above each image.}
    \label{fig:gpSequence}
\end{figure*}

\begin{figure}
    \centering
    \begin{minipage}{0.33\linewidth}
        \centering
        \scriptsize{(a) NRMSE = 0.183}\\
        \includegraphics[width=0.95\linewidth]{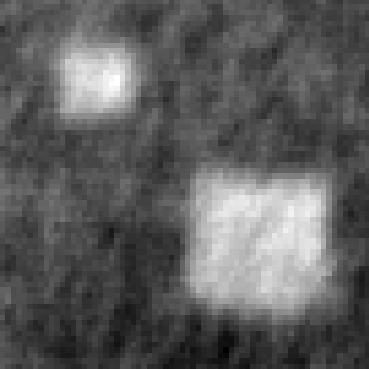}
    \end{minipage}%
    \begin{minipage}{0.33\linewidth}
        \centering
        \scriptsize{(b) NRMSE = 0.112}\\
        \includegraphics[width=0.95\linewidth]{./figures/twoUptwoDown_GP.png}
    \end{minipage}%
    \begin{minipage}{0.33\linewidth}
        \centering
        \scriptsize{(c) NRMSE = 0.203}\\
        \includegraphics[width=0.95\linewidth]{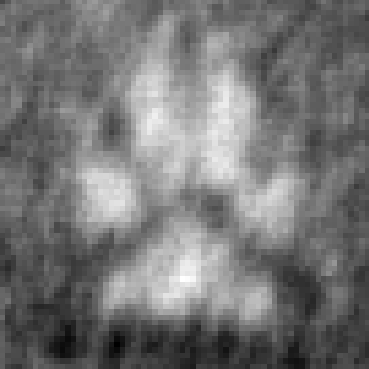}
    \end{minipage}\\
    \begin{minipage}{0.33\linewidth}
        \centering
        \scriptsize{(d) NRMSE = 0.224}\\
        \includegraphics[width=0.95\linewidth]{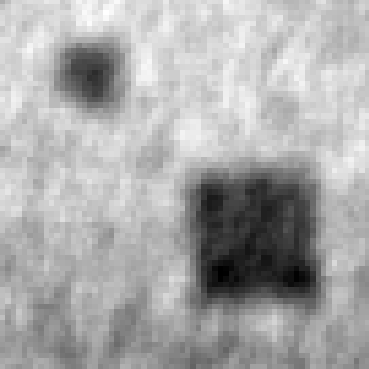}
    \end{minipage}%
    \begin{minipage}{0.33\linewidth}
        \centering
        \scriptsize{(e) NRMSE = 0.113}\\
        \includegraphics[width=0.95\linewidth]{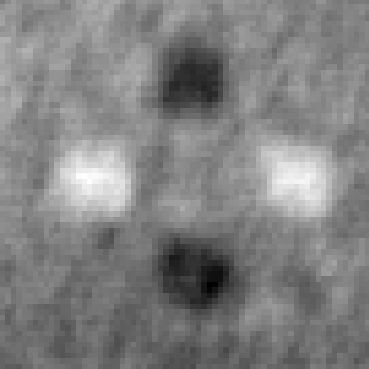}
    \end{minipage}%
    \begin{minipage}{0.33\linewidth}
        \centering
        \scriptsize{(f) NRMSE = 0.173}\\
        \includegraphics[width=0.95\linewidth]{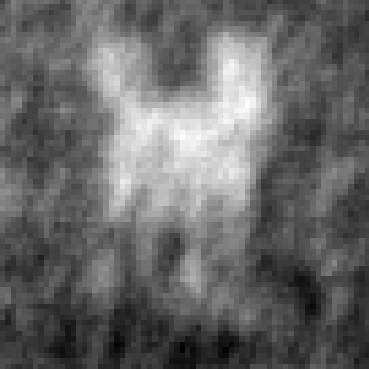}
    \end{minipage}
    \caption{Images constructed by universal-mask ghost projection with 10 grey levels. Each image corresponds to that of the target images in Fig.~\ref{fig:targetImages}, with physical dimensions 2 mm $\times$ 2 mm represented as $68 \times 68$ pixels. Corresponding videos, for each panel, are given in the Supplementary Material. Normalized RMSE (NRMSE) is reported above each image.}
    \label{fig:gpImages}
\end{figure}
\begin{figure}
    \centering
    \includegraphics[width=0.9\linewidth]{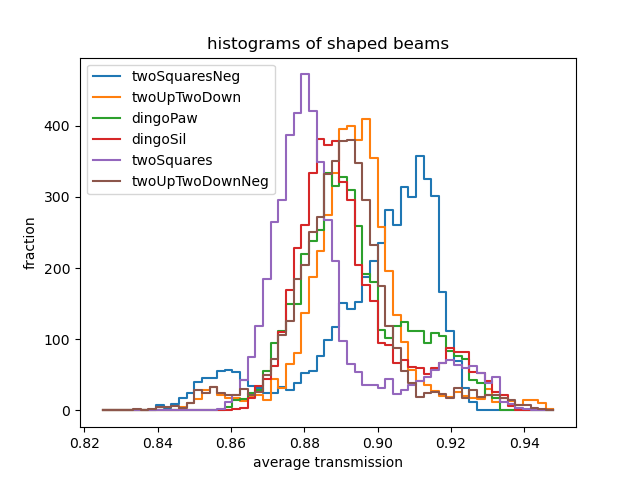}
    \caption{Histogram of ghost projection images in Fig.~\ref{fig:gpImages}. {\it twoSquares} is the large and small white square on a black background (Fig.~\ref{fig:gpImages}(a)), {\it twoSquaresNeg} is the negative of this (Fig.~\ref{fig:gpImages}(d)), {\it twoUpTwoDown} is the two white and two black squares on a grey backgound (Fig.~\ref{fig:gpImages}(b)), {\it twoUpTwoDownNeg} is the negative of this (Fig.~\ref{fig:gpImages}(e)), {\it dingoPaw} is the white dingo paw print (Fig.~\ref{fig:gpImages}(c)), and {\it dingoSil} is the silhouette of a dingo (Fig.~\ref{fig:gpImages}(f)).}
    \label{fig:gpHistograms}
\end{figure}

%
The six projected neutron beam shapes corresponding to the target images in Fig.~\ref{fig:targetImages}  are presented in Fig.~\ref{fig:gpImages}. The conservative estimate of resolution of each image determined by the 1-bit threshold in Fourier ring correlation was around 320 $\mu$m. The resolutions in micrometers corresponding to each image are as follows (a) 310 (b) 350 (c) 290 (d) 320 (e) 320 (f) 330. Normalized RMSE (NRMSE) is reported above each image in Fig. \ref{fig:gpImages}. A histogram of the transmission levels in each projected image is given in Fig.~\ref{fig:gpHistograms}.

\section{Discussion}
\label{sec:discussion}

Beam shaping of the six target images, demonstrating simple and more complex features with both positive and negative contrast, has been achieved. The images were each 2 mm $\times$ 2 mm in dimension and were, on average, constructed from 170 unique patterns of the 2307 basis set with varying exposure time. On average the images produced had a resolution of 320 $\mu$m and a normalized RMSE of 0.17. The results in Fig. \ref{fig:gpImages}, while satisfactory, are not as high quality as demonstrated by simulation in Sec. \ref{sec:simulations} using the same mask patterns. This was due to experimental factors such as shot noise in neutron detection with limited flux, penumbral blurring from pattern-to-detector distance, and the reduced pattern basis missing 193 patterns due to data transfer failure. The designed patterns used in simulation had a resolution of 20 $\mu$m while the patterns when measured experimentally only had a resolution of 140 $\mu$m. Observe in Fig. \ref{fig:pedestalVaryMetrics}b for the simulations that ghost projections with a resolution of 32 $\mu$m were achievable; $1.6\times$ worse resolution than the patterns. Experimentally, we achieved projections with 320 $\mu$m giving a $2.3\times$ degradation.

Here we maximized image fidelity possible (given the extremely limited basis set) adding a constant offset, or pedestal, to the images. The histograms in Fig. \ref{fig:gpHistograms} showed the images had an average pedestal of 0.89 with the signal standard deviation of 0.16. The pedestal could be reduced by sacrificing image quality as described in Sec. \ref{sec:simulations}. The main limitation to image quality in this experimental demonstration (and by design the simulations as well) was the modest basis set that was far from ``overcomplete'' given the available experiment time.

While a highly overcomplete \cite{MandelWolf} nonorthogonal basis is in general less efficient than a complete orthogonal basis for the purposes of synthesizing an {\em arbitrary} function \cite{Gorban2016, Gureyev2018}, for synthesizing {\em any one particular function} (namely the desired ghost projection), one can judiciously discard most members of the overcomplete set (in a manner that depends on the desired ghost projection) to obtain an efficient scheme that has fewer nonzero weighting coefficients than would be the case for a complete orthogonal basis \cite{ceddia2022ghostI, ceddia2022ghostII, ceddia2023universal, Monro2024}. In this context, observe that the overcomplete nature of the basis-pattern set implies the vector, $w$, of nonnegative weighting coefficients is highly non-unique, with the degree of nonuniqueness becoming progressively larger as the number of candidate basis patterns $K$ becomes arbitrarily large.  This nonuniqueness is an advantage because one may choose a particular weighting vector, $w$, using the criterion of sparseness, namely the demand that the number of nonzero elements of $w$ be minimized. In this sense, use of an overcomplete random-pattern basis can be viewed as a reversed form of compressed sensing \cite{CandesTao2006, Donoho2006, Rani2017}, as it uses compressive-sensing concepts for the purpose of image synthesis via ghost projection, rather than image measurement and decomposition \cite{ceddia2022ghostI, ceddia2022ghostII, ceddia2023universal, Monro2024}. Rather unusually, image compression occurs prior to image formation, via the choice of a different sparse-basis subset of the overcomplete basis for the purposes of creating each particular ghost projection.

\section{Conclusions and future work}
\label{sec:conclusion}

The ghost-projection approach to neutron beam shaping reported in this paper enables an arbitrary time-integrated intensity distribution to be created over any imaging surface. The method differs from conventional strategies that seek to achieve arbitrary-profile matter-wave beam shaping, in the sense that {\em single-shot matter-wave beam-shaping with a dynamically configurable mask is replaced by multi-shot beam-shaping using a single random mask} which is displaced to a number of transverse positions during the exposure process. Since random patterns are thereby added to make any desired image, the method can be thought of as ``building signals out of noise''.

We have presented a numerical method that can be tuned to select the patterns depending on whether image fidelity is preferable (with the sacrifice of a constant offset, or pedestal) or minimizing the dose is of utmost importance. Using this method, and opting for image fidelity, we have demonstrated neutron beam shaping of six target patterns displaying varying degrees of complexity along with positive and negative contrast.

Some future research avenues are as follows: 

(i) The ghost-projection principle is applicable to a variety of matter-wave and radiation fields. This may be of particular interest in contexts where no on-demand beam-shaping methods exist that have an appreciable degree of spatial resolution.  For example, our ghost-projection beam-shaping method could be applied to cold atom beams \cite{AtomBeamBook, AtomLasers2013}, with masks generated by atom refraction through the ponderomotive potential \cite{Kapitza_Dirac_1933, Mcclelland1993, Freimund2001} produced by an ensemble of highly-structured laser-field light sheets.
    
(ii) A continuous-exposure GP scheme is possible, whereby a single random mask is transversely displaced in a continuous manner, such that the integral over time $t$---of the resulting time-dependent intensity distribution $I(x,y,t)$ over the target image plane---generates a desired ghost projection \cite{paganin2019writing}.  Thus, each ghost projection may be encoded in a given mask-displacement trajectory $(\Delta x(t), \Delta y(t))$. This enables {\em space-time matter-wave shaping}, namely beam shaping in $(2+1)$-dimensions, if one suitably coarse grains (averages) over the time variable (by the time scale $T$ used to form each frame of the ``ghost projection movie'').
    
(iii) While we have restricted consideration to random-fractal \cite{SethnaBook} speckled masks, the ghost-projection method is more general, with the only requirement being reproducibly able to make a large set of different highly-structured patterns over the area of the ghost-projection field of view. For example, the ghost-projection concept may also be used for an ensemble of diffraction catastrophes \cite{kravtsov2012caustics} produced in the focal plane of a highly aberrated coherently-illuminated lens \cite{Monro2024}. In turn, such a focused-beam approach enables miniaturization to be built into the ghost-projection scheme, e.g.~in the electron-microscopy or hard-x-ray domains.

(iv) A spectral version of the method is also possible, whereby an ensemble of polyenergetic speckle fields may be used to create a GP movie where a desired energy spectrum is created in each GP frame. While this ``color ghost projection'' concept has been demonstrated in principle via theory and simulation \cite{ceddia2022ghostI}, it has yet to be achieved in experiment.

(v) Ghost projection onto curved rather than planar surfaces is conceptually straightforward \cite{ceddia2022ghostII}.

(vi) The number $K$ of candidate random-mask patterns can be made exponentially larger,
via two or more random masks in series, with each mask on an independent transverse translation stage \cite{ceddia2022ghostI, ceddia2022ghostII}.  If masks and source are sufficiently well characterized, each possible mask pattern may be calculated rather than measured. This enables the number of nonzero weights in $w$ to be reduced still further, compared to single-mask schemes. 

\begin{acknowledgments}
AMK and DMP thank the Australian Research Council (ARC) for funding through the Discovery Project: DP210101312. AMK and UG thank the ARC and industry partners funding the Industrial Transformation and Training Centre for Multiscale 3D Imaging, Modelling, and Manufacturing: IC180100008. Experiments were carried out under the merit proposals P16514 and P18844 at the Dingo neutron imaging beamline at the ANSTO OPAL reactor, Lucas Heights, Australia. LR thanks the ARC for funding through the Discovery Early Career Researcher Award DE240100006. We thank Dr Frank Darmann, Ravi Pushparaj, Eno Imamovic, Jason Christoforidis, Ferdi Franceschini, Jeremy Shalala, Adam Gallaty, and John Affleck for the design, fabrication, assembly, installation, and control software of the mask manipulation stages.
\end{acknowledgments}


\bibliography{references}

\begin{thebibliography}{57}%
\makeatletter
\providecommand \@ifxundefined [1]{%
 \@ifx{#1\undefined}
}%
\providecommand \@ifnum [1]{%
 \ifnum #1\expandafter \@firstoftwo
 \else \expandafter \@secondoftwo
 \fi
}%
\providecommand \@ifx [1]{%
 \ifx #1\expandafter \@firstoftwo
 \else \expandafter \@secondoftwo
 \fi
}%
\providecommand \natexlab [1]{#1}%
\providecommand \enquote  [1]{``#1''}%
\providecommand \bibnamefont  [1]{#1}%
\providecommand \bibfnamefont [1]{#1}%
\providecommand \citenamefont [1]{#1}%
\providecommand \href@noop [0]{\@secondoftwo}%
\providecommand \href [0]{\begingroup \@sanitize@url \@href}%
\providecommand \@href[1]{\@@startlink{#1}\@@href}%
\providecommand \@@href[1]{\endgroup#1\@@endlink}%
\providecommand \@sanitize@url [0]{\catcode `\\12\catcode `\$12\catcode `\&12\catcode `\#12\catcode `\^12\catcode `\_12\catcode `\%12\relax}%
\providecommand \@@startlink[1]{}%
\providecommand \@@endlink[0]{}%
\providecommand \url  [0]{\begingroup\@sanitize@url \@url }%
\providecommand \@url [1]{\endgroup\@href {#1}{\urlprefix }}%
\providecommand \urlprefix  [0]{URL }%
\providecommand \Eprint [0]{\href }%
\providecommand \doibase [0]{https://doi.org/}%
\providecommand \selectlanguage [0]{\@gobble}%
\providecommand \bibinfo  [0]{\@secondoftwo}%
\providecommand \bibfield  [0]{\@secondoftwo}%
\providecommand \translation [1]{[#1]}%
\providecommand \BibitemOpen [0]{}%
\providecommand \bibitemStop [0]{}%
\providecommand \bibitemNoStop [0]{.\EOS\space}%
\providecommand \EOS [0]{\spacefactor3000\relax}%
\providecommand \BibitemShut  [1]{\csname bibitem#1\endcsname}%
\let\auto@bib@innerbib\@empty
\bibitem [{\citenamefont {Behling}\ and\ \citenamefont {Gr{\"u}ner}(2018)}]{behling2018diagnostic}%
  \BibitemOpen
  \bibfield  {author} {\bibinfo {author} {\bibfnamefont {R.}~\bibnamefont {Behling}}\ and\ \bibinfo {author} {\bibfnamefont {F.}~\bibnamefont {Gr{\"u}ner}},\ }\bibfield  {title} {\bibinfo {title} {Diagnostic x-ray sources—present and future},\ }\href@noop {} {\bibfield  {journal} {\bibinfo  {journal} {Nucl. Instrum. Methods Phys. Res. A}\ }\textbf {\bibinfo {volume} {878}},\ \bibinfo {pages} {50} (\bibinfo {year} {2018})}\BibitemShut {NoStop}%
\bibitem [{\citenamefont {Greene}\ and\ \citenamefont {Williams}(2017)}]{greene2017linear}%
  \BibitemOpen
  \bibfield  {author} {\bibinfo {author} {\bibfnamefont {D.}~\bibnamefont {Greene}}\ and\ \bibinfo {author} {\bibfnamefont {P.~C.}\ \bibnamefont {Williams}},\ }\href@noop {} {\emph {\bibinfo {title} {Linear Accelerators for Radiation Therapy}}}\ (\bibinfo  {publisher} {CRC Press},\ \bibinfo {year} {2017})\BibitemShut {NoStop}%
\bibitem [{\citenamefont {Degiovanni}\ and\ \citenamefont {Amaldi}(2015)}]{degiovanni2015history}%
  \BibitemOpen
  \bibfield  {author} {\bibinfo {author} {\bibfnamefont {A.}~\bibnamefont {Degiovanni}}\ and\ \bibinfo {author} {\bibfnamefont {U.}~\bibnamefont {Amaldi}},\ }\bibfield  {title} {\bibinfo {title} {History of hadron therapy accelerators},\ }\href@noop {} {\bibfield  {journal} {\bibinfo  {journal} {Phys. Med.}\ }\textbf {\bibinfo {volume} {31}},\ \bibinfo {pages} {322} (\bibinfo {year} {2015})}\BibitemShut {NoStop}%
\bibitem [{\citenamefont {Crowder}(2013)}]{crowder2013ion}%
  \BibitemOpen
  \bibfield  {author} {\bibinfo {author} {\bibfnamefont {B.}~\bibnamefont {Crowder}},\ }\href@noop {} {\emph {\bibinfo {title} {Ion Implantation in Semiconductors and other Materials}}}\ (\bibinfo  {publisher} {Springer Science \& Business Media},\ \bibinfo {year} {2013})\BibitemShut {NoStop}%
\bibitem [{\citenamefont {Wang}\ \emph {et~al.}(2020)\citenamefont {Wang}, \citenamefont {Zhong}, \citenamefont {Lee}, \citenamefont {Fancey},\ and\ \citenamefont {Mi}}]{wang2020non}%
  \BibitemOpen
  \bibfield  {author} {\bibinfo {author} {\bibfnamefont {B.}~\bibnamefont {Wang}}, \bibinfo {author} {\bibfnamefont {S.}~\bibnamefont {Zhong}}, \bibinfo {author} {\bibfnamefont {T.-L.}\ \bibnamefont {Lee}}, \bibinfo {author} {\bibfnamefont {K.~S.}\ \bibnamefont {Fancey}},\ and\ \bibinfo {author} {\bibfnamefont {J.}~\bibnamefont {Mi}},\ }\bibfield  {title} {\bibinfo {title} {Non-destructive testing and evaluation of composite materials/structures: A state-of-the-art review},\ }\href@noop {} {\bibfield  {journal} {\bibinfo  {journal} {Adv. Mech. Eng.}\ }\textbf {\bibinfo {volume} {12}},\ \bibinfo {pages} {1687814020913761} (\bibinfo {year} {2020})}\BibitemShut {NoStop}%
\bibitem [{\citenamefont {Wells}\ and\ \citenamefont {Bradley}(2012)}]{wells2012review}%
  \BibitemOpen
  \bibfield  {author} {\bibinfo {author} {\bibfnamefont {K.}~\bibnamefont {Wells}}\ and\ \bibinfo {author} {\bibfnamefont {D.}~\bibnamefont {Bradley}},\ }\bibfield  {title} {\bibinfo {title} {A review of x-ray explosives detection techniques for checked baggage},\ }\href@noop {} {\bibfield  {journal} {\bibinfo  {journal} {Appl. Radiat. Isot.}\ }\textbf {\bibinfo {volume} {70}},\ \bibinfo {pages} {1729} (\bibinfo {year} {2012})}\BibitemShut {NoStop}%
\bibitem [{\citenamefont {Akcay}\ and\ \citenamefont {Breckon}(2022)}]{akcay2022towards}%
  \BibitemOpen
  \bibfield  {author} {\bibinfo {author} {\bibfnamefont {S.}~\bibnamefont {Akcay}}\ and\ \bibinfo {author} {\bibfnamefont {T.}~\bibnamefont {Breckon}},\ }\bibfield  {title} {\bibinfo {title} {Towards automatic threat detection: A survey of advances of deep learning within x-ray security imaging},\ }\href@noop {} {\bibfield  {journal} {\bibinfo  {journal} {Pattern Recognit.}\ }\textbf {\bibinfo {volume} {122}},\ \bibinfo {pages} {108245} (\bibinfo {year} {2022})}\BibitemShut {NoStop}%
\bibitem [{\citenamefont {Hanke}\ \emph {et~al.}(2008)\citenamefont {Hanke}, \citenamefont {Fuchs},\ and\ \citenamefont {Uhlmann}}]{hanke2008x}%
  \BibitemOpen
  \bibfield  {author} {\bibinfo {author} {\bibfnamefont {R.}~\bibnamefont {Hanke}}, \bibinfo {author} {\bibfnamefont {T.}~\bibnamefont {Fuchs}},\ and\ \bibinfo {author} {\bibfnamefont {N.}~\bibnamefont {Uhlmann}},\ }\bibfield  {title} {\bibinfo {title} {X-ray based methods for non-destructive testing and material characterization},\ }\href@noop {} {\bibfield  {journal} {\bibinfo  {journal} {Nucl. Instrum. Methods Phys. Res. A}\ }\textbf {\bibinfo {volume} {591}},\ \bibinfo {pages} {14} (\bibinfo {year} {2008})}\BibitemShut {NoStop}%
\bibitem [{\citenamefont {Towsyfyan}\ \emph {et~al.}(2020)\citenamefont {Towsyfyan}, \citenamefont {Biguri}, \citenamefont {Boardman},\ and\ \citenamefont {Blumensath}}]{towsyfyan2020successes}%
  \BibitemOpen
  \bibfield  {author} {\bibinfo {author} {\bibfnamefont {H.}~\bibnamefont {Towsyfyan}}, \bibinfo {author} {\bibfnamefont {A.}~\bibnamefont {Biguri}}, \bibinfo {author} {\bibfnamefont {R.}~\bibnamefont {Boardman}},\ and\ \bibinfo {author} {\bibfnamefont {T.}~\bibnamefont {Blumensath}},\ }\bibfield  {title} {\bibinfo {title} {Successes and challenges in non-destructive testing of aircraft composite structures},\ }\href@noop {} {\bibfield  {journal} {\bibinfo  {journal} {Chinese J. Aeronaut.}\ }\textbf {\bibinfo {volume} {33}},\ \bibinfo {pages} {771} (\bibinfo {year} {2020})}\BibitemShut {NoStop}%
\bibitem [{\citenamefont {Klein}\ and\ \citenamefont {Werner}(1983)}]{KleinWerner1983}%
  \BibitemOpen
  \bibfield  {author} {\bibinfo {author} {\bibfnamefont {A.~G.}\ \bibnamefont {Klein}}\ and\ \bibinfo {author} {\bibfnamefont {S.~A.}\ \bibnamefont {Werner}},\ }\bibfield  {title} {\bibinfo {title} {Neutron optics},\ }\href {https://doi.org/10.1088/0034-4885/46/3/001} {\bibfield  {journal} {\bibinfo  {journal} {Rep. Prog. Phys.}\ }\textbf {\bibinfo {volume} {46}},\ \bibinfo {pages} {259} (\bibinfo {year} {1983})}\BibitemShut {NoStop}%
\bibitem [{\citenamefont {Eskildsen}\ \emph {et~al.}(1998)\citenamefont {Eskildsen}, \citenamefont {Gammel}, \citenamefont {Isaacs}, \citenamefont {Detlefs}, \citenamefont {Mortensen},\ and\ \citenamefont {Bishop}}]{NeutronCRL}%
  \BibitemOpen
  \bibfield  {author} {\bibinfo {author} {\bibfnamefont {M.~R.}\ \bibnamefont {Eskildsen}}, \bibinfo {author} {\bibfnamefont {P.~L.}\ \bibnamefont {Gammel}}, \bibinfo {author} {\bibfnamefont {E.~D.}\ \bibnamefont {Isaacs}}, \bibinfo {author} {\bibfnamefont {C.}~\bibnamefont {Detlefs}}, \bibinfo {author} {\bibfnamefont {K.}~\bibnamefont {Mortensen}},\ and\ \bibinfo {author} {\bibfnamefont {D.~J.}\ \bibnamefont {Bishop}},\ }\bibfield  {title} {\bibinfo {title} {Compound refractive optics for the imaging and focusing of low-energy neutrons},\ }\href@noop {} {\bibfield  {journal} {\bibinfo  {journal} {Nature}\ }\textbf {\bibinfo {volume} {391}},\ \bibinfo {pages} {563} (\bibinfo {year} {1998})}\BibitemShut {NoStop}%
\bibitem [{\citenamefont {Leemreize}\ \emph {et~al.}(2019)\citenamefont {Leemreize}, \citenamefont {Knudsen}, \citenamefont {Birk}, \citenamefont {Strobl}, \citenamefont {Detlefs},\ and\ \citenamefont {Poulsen}}]{NeutronCRL2}%
  \BibitemOpen
  \bibfield  {author} {\bibinfo {author} {\bibfnamefont {H.}~\bibnamefont {Leemreize}}, \bibinfo {author} {\bibfnamefont {E.~B.}\ \bibnamefont {Knudsen}}, \bibinfo {author} {\bibfnamefont {J.~O.}\ \bibnamefont {Birk}}, \bibinfo {author} {\bibfnamefont {M.}~\bibnamefont {Strobl}}, \bibinfo {author} {\bibfnamefont {C.}~\bibnamefont {Detlefs}},\ and\ \bibinfo {author} {\bibfnamefont {H.~F.}\ \bibnamefont {Poulsen}},\ }\bibfield  {title} {\bibinfo {title} {{Full-field neutron microscopy based on refractive optics}},\ }\href {https://doi.org/10.1107/S1600576719012858} {\bibfield  {journal} {\bibinfo  {journal} {J. Appl. Cryst.}\ }\textbf {\bibinfo {volume} {52}},\ \bibinfo {pages} {1299} (\bibinfo {year} {2019})}\BibitemShut {NoStop}%
\bibitem [{\citenamefont {Reid}(2008)}]{DataProjectorReference}%
  \BibitemOpen
  \bibfield  {author} {\bibinfo {author} {\bibfnamefont {A.}~\bibnamefont {Reid}},\ }\bibfield  {title} {\bibinfo {title} {The physics of the data projector},\ }\href {https://doi.org/10.1088/0031-9120/43/6/006} {\bibfield  {journal} {\bibinfo  {journal} {Phys. Educ.}\ }\textbf {\bibinfo {volume} {43}},\ \bibinfo {pages} {599} (\bibinfo {year} {2008})}\BibitemShut {NoStop}%
\bibitem [{\citenamefont {Zhu}\ and\ \citenamefont {Wang}(2014)}]{SLM-reference}%
  \BibitemOpen
  \bibfield  {author} {\bibinfo {author} {\bibfnamefont {L.}~\bibnamefont {Zhu}}\ and\ \bibinfo {author} {\bibfnamefont {J.}~\bibnamefont {Wang}},\ }\bibfield  {title} {\bibinfo {title} {Arbitrary manipulation of spatial amplitude and phase using phase-only spatial light modulators},\ }\href@noop {} {\bibfield  {journal} {\bibinfo  {journal} {Sci. Rep.}\ }\textbf {\bibinfo {volume} {4}},\ \bibinfo {pages} {7441} (\bibinfo {year} {2014})}\BibitemShut {NoStop}%
\bibitem [{\citenamefont {Gr\"{u}nauer}(2005)}]{Grunauer2005}%
  \BibitemOpen
  \bibfield  {author} {\bibinfo {author} {\bibfnamefont {F.}~\bibnamefont {Gr\"{u}nauer}},\ }\bibfield  {title} {\bibinfo {title} {Image deconvolution and coded masks in neutron radiography},\ }\href {https://doi.org/https://doi.org/10.1016/j.nima.2005.01.160} {\bibfield  {journal} {\bibinfo  {journal} {Nucl. Instrum. Methods Phys. Res. A}\ }\textbf {\bibinfo {volume} {542}},\ \bibinfo {pages} {342} (\bibinfo {year} {2005})}\BibitemShut {NoStop}%
\bibitem [{\citenamefont {Tamasaku}\ \emph {et~al.}(2024)\citenamefont {Tamasaku}, \citenamefont {Sato}, \citenamefont {Osaka}, \citenamefont {Osawa}, \citenamefont {Zhu},\ and\ \citenamefont {Ishikawa}}]{Tamasaku2024}%
  \BibitemOpen
  \bibfield  {author} {\bibinfo {author} {\bibfnamefont {K.}~\bibnamefont {Tamasaku}}, \bibinfo {author} {\bibfnamefont {T.}~\bibnamefont {Sato}}, \bibinfo {author} {\bibfnamefont {T.}~\bibnamefont {Osaka}}, \bibinfo {author} {\bibfnamefont {H.}~\bibnamefont {Osawa}}, \bibinfo {author} {\bibfnamefont {D.}~\bibnamefont {Zhu}},\ and\ \bibinfo {author} {\bibfnamefont {T.}~\bibnamefont {Ishikawa}},\ }\bibfield  {title} {\bibinfo {title} {Dynamically patterning x-ray beam by a femtosecond optical laser},\ }\href {https://doi.org/10.1126/sciadv.adp5326} {\bibfield  {journal} {\bibinfo  {journal} {Sci. Adv.}\ }\textbf {\bibinfo {volume} {10}},\ \bibinfo {pages} {eadp5326} (\bibinfo {year} {2024})}\BibitemShut {NoStop}%
\bibitem [{\citenamefont {Paganin}(2019)}]{paganin2019writing}%
  \BibitemOpen
  \bibfield  {author} {\bibinfo {author} {\bibfnamefont {D.~M.}\ \bibnamefont {Paganin}},\ }\bibfield  {title} {\bibinfo {title} {Writing arbitrary distributions of radiant exposure by scanning a single illuminated spatially random screen},\ }\href@noop {} {\bibfield  {journal} {\bibinfo  {journal} {Phys. Rev. A}\ }\textbf {\bibinfo {volume} {100}},\ \bibinfo {pages} {063823} (\bibinfo {year} {2019})}\BibitemShut {NoStop}%
\bibitem [{\citenamefont {Ceddia}\ and\ \citenamefont {Paganin}(2022)}]{ceddia2022ghostI}%
  \BibitemOpen
  \bibfield  {author} {\bibinfo {author} {\bibfnamefont {D.}~\bibnamefont {Ceddia}}\ and\ \bibinfo {author} {\bibfnamefont {D.~M.}\ \bibnamefont {Paganin}},\ }\bibfield  {title} {\bibinfo {title} {Ghost projection},\ }\href@noop {} {\bibfield  {journal} {\bibinfo  {journal} {Phys. Rev. A}\ }\textbf {\bibinfo {volume} {105}},\ \bibinfo {pages} {013512} (\bibinfo {year} {2022})}\BibitemShut {NoStop}%
\bibitem [{\citenamefont {Ceddia}\ \emph {et~al.}(2022)\citenamefont {Ceddia}, \citenamefont {Kingston}, \citenamefont {Pelliccia}, \citenamefont {Rack},\ and\ \citenamefont {Paganin}}]{ceddia2022ghostII}%
  \BibitemOpen
  \bibfield  {author} {\bibinfo {author} {\bibfnamefont {D.}~\bibnamefont {Ceddia}}, \bibinfo {author} {\bibfnamefont {A.~M.}\ \bibnamefont {Kingston}}, \bibinfo {author} {\bibfnamefont {D.}~\bibnamefont {Pelliccia}}, \bibinfo {author} {\bibfnamefont {A.}~\bibnamefont {Rack}},\ and\ \bibinfo {author} {\bibfnamefont {D.~M.}\ \bibnamefont {Paganin}},\ }\bibfield  {title} {\bibinfo {title} {Ghost projection. {II}. {B}eam shaping using realistic spatially random masks},\ }\href@noop {} {\bibfield  {journal} {\bibinfo  {journal} {Phys. Rev. A}\ }\textbf {\bibinfo {volume} {106}},\ \bibinfo {pages} {033512} (\bibinfo {year} {2022})}\BibitemShut {NoStop}%
\bibitem [{\citenamefont {Shapiro}(2008)}]{shapiro2008computational}%
  \BibitemOpen
  \bibfield  {author} {\bibinfo {author} {\bibfnamefont {J.~H.}\ \bibnamefont {Shapiro}},\ }\bibfield  {title} {\bibinfo {title} {Computational ghost imaging},\ }\href@noop {} {\bibfield  {journal} {\bibinfo  {journal} {Phys. Rev. A}\ }\textbf {\bibinfo {volume} {78}},\ \bibinfo {pages} {061802} (\bibinfo {year} {2008})}\BibitemShut {NoStop}%
\bibitem [{\citenamefont {Erkmen}\ and\ \citenamefont {Shapiro}(2010)}]{erkmen2010ghost}%
  \BibitemOpen
  \bibfield  {author} {\bibinfo {author} {\bibfnamefont {B.~I.}\ \bibnamefont {Erkmen}}\ and\ \bibinfo {author} {\bibfnamefont {J.~H.}\ \bibnamefont {Shapiro}},\ }\bibfield  {title} {\bibinfo {title} {Ghost imaging: from quantum to classical to computational},\ }\href@noop {} {\bibfield  {journal} {\bibinfo  {journal} {Adv. Opt. Photonics}\ }\textbf {\bibinfo {volume} {2}},\ \bibinfo {pages} {405} (\bibinfo {year} {2010})}\BibitemShut {NoStop}%
\bibitem [{\citenamefont {Shapiro}\ and\ \citenamefont {Boyd}(2012)}]{Shapiro2012}%
  \BibitemOpen
  \bibfield  {author} {\bibinfo {author} {\bibfnamefont {J.~H.}\ \bibnamefont {Shapiro}}\ and\ \bibinfo {author} {\bibfnamefont {R.~W.}\ \bibnamefont {Boyd}},\ }\bibfield  {title} {\bibinfo {title} {The physics of ghost imaging},\ }\href@noop {} {\bibfield  {journal} {\bibinfo  {journal} {Quantum Inf. Process.}\ }\textbf {\bibinfo {volume} {11}},\ \bibinfo {pages} {949} (\bibinfo {year} {2012})}\BibitemShut {NoStop}%
\bibitem [{\citenamefont {Padgett}\ and\ \citenamefont {Boyd}(2017)}]{Padgett2017}%
  \BibitemOpen
  \bibfield  {author} {\bibinfo {author} {\bibfnamefont {M.~J.}\ \bibnamefont {Padgett}}\ and\ \bibinfo {author} {\bibfnamefont {R.~W.}\ \bibnamefont {Boyd}},\ }\bibfield  {title} {\bibinfo {title} {An introduction to ghost imaging: quantum and classical},\ }\href@noop {} {\bibfield  {journal} {\bibinfo  {journal} {Phil. Trans. R. Soc. A}\ }\textbf {\bibinfo {volume} {375}},\ \bibinfo {pages} {20160233} (\bibinfo {year} {2017})}\BibitemShut {NoStop}%
\bibitem [{\citenamefont {Ceddia}\ \emph {et~al.}(2023)\citenamefont {Ceddia}, \citenamefont {Aminzadeh}, \citenamefont {Cook}, \citenamefont {Pelliccia}, \citenamefont {Kingston},\ and\ \citenamefont {Paganin}}]{ceddia2023universal}%
  \BibitemOpen
  \bibfield  {author} {\bibinfo {author} {\bibfnamefont {D.}~\bibnamefont {Ceddia}}, \bibinfo {author} {\bibfnamefont {A.}~\bibnamefont {Aminzadeh}}, \bibinfo {author} {\bibfnamefont {P.~K.}\ \bibnamefont {Cook}}, \bibinfo {author} {\bibfnamefont {D.}~\bibnamefont {Pelliccia}}, \bibinfo {author} {\bibfnamefont {A.~M.}\ \bibnamefont {Kingston}},\ and\ \bibinfo {author} {\bibfnamefont {D.~M.}\ \bibnamefont {Paganin}},\ }\bibfield  {title} {\bibinfo {title} {Universal mask for hard x rays},\ }\href@noop {} {\bibfield  {journal} {\bibinfo  {journal} {Optica}\ }\textbf {\bibinfo {volume} {10}},\ \bibinfo {pages} {1067} (\bibinfo {year} {2023})}\BibitemShut {NoStop}%
\bibitem [{\citenamefont {Kingston}\ \emph {et~al.}(2023)\citenamefont {Kingston}, \citenamefont {Aminzadeh}, \citenamefont {Roberts}, \citenamefont {Pelliccia}, \citenamefont {Svalbe},\ and\ \citenamefont {Paganin}}]{kingston2023optimizing}%
  \BibitemOpen
  \bibfield  {author} {\bibinfo {author} {\bibfnamefont {A.~M.}\ \bibnamefont {Kingston}}, \bibinfo {author} {\bibfnamefont {A.}~\bibnamefont {Aminzadeh}}, \bibinfo {author} {\bibfnamefont {L.}~\bibnamefont {Roberts}}, \bibinfo {author} {\bibfnamefont {D.}~\bibnamefont {Pelliccia}}, \bibinfo {author} {\bibfnamefont {I.~D.}\ \bibnamefont {Svalbe}},\ and\ \bibinfo {author} {\bibfnamefont {D.~M.}\ \bibnamefont {Paganin}},\ }\bibfield  {title} {\bibinfo {title} {Optimizing nonconfigurable, transversely displaced masks for illumination patterns in classical ghost imaging},\ }\href@noop {} {\bibfield  {journal} {\bibinfo  {journal} {Phys. Rev. A}\ }\textbf {\bibinfo {volume} {107}},\ \bibinfo {pages} {023524} (\bibinfo {year} {2023})}\BibitemShut {NoStop}%
\bibitem [{\citenamefont {Aminzadeh}\ \emph {et~al.}(2023)\citenamefont {Aminzadeh}, \citenamefont {Roberts}, \citenamefont {Young}, \citenamefont {Chiang}, \citenamefont {Svalbe}, \citenamefont {Paganin},\ and\ \citenamefont {Kingston}}]{aminzadeh2023mask}%
  \BibitemOpen
  \bibfield  {author} {\bibinfo {author} {\bibfnamefont {A.}~\bibnamefont {Aminzadeh}}, \bibinfo {author} {\bibfnamefont {L.}~\bibnamefont {Roberts}}, \bibinfo {author} {\bibfnamefont {B.}~\bibnamefont {Young}}, \bibinfo {author} {\bibfnamefont {C.-I.}\ \bibnamefont {Chiang}}, \bibinfo {author} {\bibfnamefont {I.~D.}\ \bibnamefont {Svalbe}}, \bibinfo {author} {\bibfnamefont {D.~M.}\ \bibnamefont {Paganin}},\ and\ \bibinfo {author} {\bibfnamefont {A.~M.}\ \bibnamefont {Kingston}},\ }\bibfield  {title} {\bibinfo {title} {Mask design, fabrication, and experimental ghost imaging applications for patterned x-ray illumination},\ }\href@noop {} {\bibfield  {journal} {\bibinfo  {journal} {Opt. Express}\ }\textbf {\bibinfo {volume} {31}},\ \bibinfo {pages} {24328} (\bibinfo {year} {2023})}\BibitemShut {NoStop}%
\bibitem [{\citenamefont {Kingston}\ \emph {et~al.}(2020)\citenamefont {Kingston}, \citenamefont {Myers}, \citenamefont {Pelliccia}, \citenamefont {Salvemini}, \citenamefont {Bevitt}, \citenamefont {Garbe},\ and\ \citenamefont {Paganin}}]{kingston2020neutron}%
  \BibitemOpen
  \bibfield  {author} {\bibinfo {author} {\bibfnamefont {A.~M.}\ \bibnamefont {Kingston}}, \bibinfo {author} {\bibfnamefont {G.~R.}\ \bibnamefont {Myers}}, \bibinfo {author} {\bibfnamefont {D.}~\bibnamefont {Pelliccia}}, \bibinfo {author} {\bibfnamefont {F.}~\bibnamefont {Salvemini}}, \bibinfo {author} {\bibfnamefont {J.~J.}\ \bibnamefont {Bevitt}}, \bibinfo {author} {\bibfnamefont {U.}~\bibnamefont {Garbe}},\ and\ \bibinfo {author} {\bibfnamefont {D.~M.}\ \bibnamefont {Paganin}},\ }\bibfield  {title} {\bibinfo {title} {Neutron ghost imaging},\ }\href@noop {} {\bibfield  {journal} {\bibinfo  {journal} {Phys. Rev. A}\ }\textbf {\bibinfo {volume} {101}},\ \bibinfo {pages} {053844} (\bibinfo {year} {2020})}\BibitemShut {NoStop}%
\bibitem [{\citenamefont {He}\ \emph {et~al.}(2021)\citenamefont {He}, \citenamefont {Huang}, \citenamefont {Zeng}, \citenamefont {Li}, \citenamefont {Tan}, \citenamefont {Chen}, \citenamefont {Wu}, \citenamefont {Li}, \citenamefont {Quan}, \citenamefont {Wang} \emph {et~al.}}]{he2021single}%
  \BibitemOpen
  \bibfield  {author} {\bibinfo {author} {\bibfnamefont {Y.-H.}\ \bibnamefont {He}}, \bibinfo {author} {\bibfnamefont {Y.-Y.}\ \bibnamefont {Huang}}, \bibinfo {author} {\bibfnamefont {Z.-R.}\ \bibnamefont {Zeng}}, \bibinfo {author} {\bibfnamefont {Y.-F.}\ \bibnamefont {Li}}, \bibinfo {author} {\bibfnamefont {J.-H.}\ \bibnamefont {Tan}}, \bibinfo {author} {\bibfnamefont {L.-M.}\ \bibnamefont {Chen}}, \bibinfo {author} {\bibfnamefont {L.-A.}\ \bibnamefont {Wu}}, \bibinfo {author} {\bibfnamefont {M.-F.}\ \bibnamefont {Li}}, \bibinfo {author} {\bibfnamefont {B.-G.}\ \bibnamefont {Quan}}, \bibinfo {author} {\bibfnamefont {S.-L.}\ \bibnamefont {Wang}}, \emph {et~al.},\ }\bibfield  {title} {\bibinfo {title} {Single-pixel imaging with neutrons},\ }\href@noop {} {\bibfield  {journal} {\bibinfo  {journal} {Sci. Bull.}\ }\textbf {\bibinfo {volume} {66}},\ \bibinfo {pages} {133} (\bibinfo {year} {2021})}\BibitemShut {NoStop}%
\bibitem [{\citenamefont {Fomin}\ \emph {et~al.}(2022)\citenamefont {Fomin}, \citenamefont {Fry}, \citenamefont {Pattie},\ and\ \citenamefont {Greene}}]{SpallationSources}%
  \BibitemOpen
  \bibfield  {author} {\bibinfo {author} {\bibfnamefont {N.}~\bibnamefont {Fomin}}, \bibinfo {author} {\bibfnamefont {J.}~\bibnamefont {Fry}}, \bibinfo {author} {\bibfnamefont {R.~W.}\ \bibnamefont {Pattie}},\ and\ \bibinfo {author} {\bibfnamefont {G.~L.}\ \bibnamefont {Greene}},\ }\bibfield  {title} {\bibinfo {title} {Fundamental neutron physics at spallation sources},\ }\href {https://doi.org/https://doi.org/10.1146/annurev-nucl-121521-051029} {\bibfield  {journal} {\bibinfo  {journal} {Annu. Rev. Nucl. Part. Sci.}\ }\textbf {\bibinfo {volume} {72}},\ \bibinfo {pages} {151} (\bibinfo {year} {2022})}\BibitemShut {NoStop}%
\bibitem [{\citenamefont {Mandel}\ and\ \citenamefont {Wolf}(1995)}]{MandelWolf}%
  \BibitemOpen
  \bibfield  {author} {\bibinfo {author} {\bibfnamefont {L.}~\bibnamefont {Mandel}}\ and\ \bibinfo {author} {\bibfnamefont {E.}~\bibnamefont {Wolf}},\ }\href@noop {} {\emph {\bibinfo {title} {Optical Coherence and Quantum Optics}}}\ (\bibinfo  {publisher} {Cambridge University Press, Cambridge},\ \bibinfo {year} {1995})\BibitemShut {NoStop}%
\bibitem [{\citenamefont {Gould}\ \emph {et~al.}(2003)\citenamefont {Gould}, \citenamefont {Orban},\ and\ \citenamefont {Toint}}]{gould2003galahad}%
  \BibitemOpen
  \bibfield  {author} {\bibinfo {author} {\bibfnamefont {N.~I.~M.}\ \bibnamefont {Gould}}, \bibinfo {author} {\bibfnamefont {D.}~\bibnamefont {Orban}},\ and\ \bibinfo {author} {\bibfnamefont {P.~L.}\ \bibnamefont {Toint}},\ }\bibfield  {title} {\bibinfo {title} {{GALAHAD}, a library of thread-safe {F}ortran 90 packages for large-scale nonlinear optimization},\ }\href {https://doi.org/10.1145/962437.962438} {\bibfield  {journal} {\bibinfo  {journal} {ACM Transactions on Mathematical Software}\ }\textbf {\bibinfo {volume} {29}},\ \bibinfo {pages} {353} (\bibinfo {year} {2003})}\BibitemShut {NoStop}%
\bibitem [{\citenamefont {Saxton}\ and\ \citenamefont {Baumeister}(1982)}]{saxton1982correlation}%
  \BibitemOpen
  \bibfield  {author} {\bibinfo {author} {\bibfnamefont {W.~O.}\ \bibnamefont {Saxton}}\ and\ \bibinfo {author} {\bibfnamefont {W.}~\bibnamefont {Baumeister}},\ }\bibfield  {title} {\bibinfo {title} {The correlation averaging of a regularly arranged bacterial cell envelope protein},\ }\href@noop {} {\bibfield  {journal} {\bibinfo  {journal} {J. Microsc.}\ }\textbf {\bibinfo {volume} {127}},\ \bibinfo {pages} {127} (\bibinfo {year} {1982})}\BibitemShut {NoStop}%
\bibitem [{\citenamefont {Van~Heel}\ \emph {et~al.}(1982)\citenamefont {Van~Heel}, \citenamefont {Keegstra}, \citenamefont {Schutter},\ and\ \citenamefont {Van~Bruggen}}]{vanHeel1982arthropod}%
  \BibitemOpen
  \bibfield  {author} {\bibinfo {author} {\bibfnamefont {M.}~\bibnamefont {Van~Heel}}, \bibinfo {author} {\bibfnamefont {W.}~\bibnamefont {Keegstra}}, \bibinfo {author} {\bibfnamefont {W.}~\bibnamefont {Schutter}},\ and\ \bibinfo {author} {\bibfnamefont {E.}~\bibnamefont {Van~Bruggen}},\ }\bibfield  {title} {\bibinfo {title} {Arthropod hemocyanin structures studied by image analysis},\ }\href@noop {} {\bibfield  {journal} {\bibinfo  {journal} {Life Chem. Rep. Suppl}\ }\textbf {\bibinfo {volume} {1}},\ \bibinfo {pages} {5} (\bibinfo {year} {1982})}\BibitemShut {NoStop}%
\bibitem [{\citenamefont {Van~Heel}\ and\ \citenamefont {Schatz}(2005)}]{van2005fourier}%
  \BibitemOpen
  \bibfield  {author} {\bibinfo {author} {\bibfnamefont {M.}~\bibnamefont {Van~Heel}}\ and\ \bibinfo {author} {\bibfnamefont {M.}~\bibnamefont {Schatz}},\ }\bibfield  {title} {\bibinfo {title} {Fourier shell correlation threshold criteria},\ }\href@noop {} {\bibfield  {journal} {\bibinfo  {journal} {Journal of structural biology}\ }\textbf {\bibinfo {volume} {151}},\ \bibinfo {pages} {250} (\bibinfo {year} {2005})}\BibitemShut {NoStop}%
\bibitem [{\citenamefont {Garbe}\ \emph {et~al.}(2015)\citenamefont {Garbe}, \citenamefont {Randall}, \citenamefont {Hughes}, \citenamefont {Davidson}, \citenamefont {Pangelis},\ and\ \citenamefont {Kennedy}}]{Garbe2015}%
  \BibitemOpen
  \bibfield  {author} {\bibinfo {author} {\bibfnamefont {U.}~\bibnamefont {Garbe}}, \bibinfo {author} {\bibfnamefont {T.}~\bibnamefont {Randall}}, \bibinfo {author} {\bibfnamefont {C.}~\bibnamefont {Hughes}}, \bibinfo {author} {\bibfnamefont {G.}~\bibnamefont {Davidson}}, \bibinfo {author} {\bibfnamefont {S.}~\bibnamefont {Pangelis}},\ and\ \bibinfo {author} {\bibfnamefont {S.~J.}\ \bibnamefont {Kennedy}},\ }\bibfield  {title} {\bibinfo {title} {A new neutron radiography/tomography/imaging station {DINGO} at {OPAL}},\ }\href {https://doi.org/https://doi.org/10.1016/j.phpro.2015.07.003} {\bibfield  {journal} {\bibinfo  {journal} {Phys. Procedia}\ }\textbf {\bibinfo {volume} {69}},\ \bibinfo {pages} {27 } (\bibinfo {year} {2015})}\BibitemShut {NoStop}%
\bibitem [{\citenamefont {Garbe}\ \emph {et~al.}(2017)\citenamefont {Garbe}, \citenamefont {Ahuja}, \citenamefont {Ibrahim}, \citenamefont {Li}, \citenamefont {Aldridge}, \citenamefont {Salvemini},\ and\ \citenamefont {Paradowska}}]{Garbe2017}%
  \BibitemOpen
  \bibfield  {author} {\bibinfo {author} {\bibfnamefont {U.}~\bibnamefont {Garbe}}, \bibinfo {author} {\bibfnamefont {Y.}~\bibnamefont {Ahuja}}, \bibinfo {author} {\bibfnamefont {R.}~\bibnamefont {Ibrahim}}, \bibinfo {author} {\bibfnamefont {H.}~\bibnamefont {Li}}, \bibinfo {author} {\bibfnamefont {L.}~\bibnamefont {Aldridge}}, \bibinfo {author} {\bibfnamefont {F.}~\bibnamefont {Salvemini}},\ and\ \bibinfo {author} {\bibfnamefont {A.~Z.}\ \bibnamefont {Paradowska}},\ }\bibfield  {title} {\bibinfo {title} {Industrial application experiments on the neutron imaging instrument {DINGO}},\ }\href {https://doi.org/https://doi.org/10.1016/j.phpro.2017.06.001} {\bibfield  {journal} {\bibinfo  {journal} {Phys. Procedia}\ }\textbf {\bibinfo {volume} {88}},\ \bibinfo {pages} {13} (\bibinfo {year} {2017})}\BibitemShut {NoStop}%
\bibitem [{\citenamefont {Treimer}(2009)}]{Treimer2009}%
  \BibitemOpen
  \bibfield  {author} {\bibinfo {author} {\bibfnamefont {W.}~\bibnamefont {Treimer}},\ }\bibinfo {title} {Neutron tomography},\ in\ \href {https://doi.org/10.1007/978-0-387-78693-3_6} {\emph {\bibinfo {booktitle} {Neutron Imaging and Applications: A Reference for the Imaging Community}}},\ \bibinfo {editor} {edited by\ \bibinfo {editor} {\bibfnamefont {I.~S.}\ \bibnamefont {Anderson}}, \bibinfo {editor} {\bibfnamefont {R.}~\bibnamefont {McGreevy}},\ and\ \bibinfo {editor} {\bibfnamefont {H.~Z.}\ \bibnamefont {Bilheux}}}\ (\bibinfo  {publisher} {Springer US},\ \bibinfo {address} {Boston, MA},\ \bibinfo {year} {2009})\ pp.\ \bibinfo {pages} {81--108}\BibitemShut {NoStop}%
\bibitem [{\citenamefont {Jakubowski}\ \emph {et~al.}(2023)\citenamefont {Jakubowski}, \citenamefont {Chacon}, \citenamefont {Tran}, \citenamefont {Stopic}, \citenamefont {Garbe}, \citenamefont {Bevitt}, \citenamefont {Olsen}, \citenamefont {Franklin}, \citenamefont {Rosenfeld}, \citenamefont {Guatelli} \emph {et~al.}}]{jakubowski2023monte}%
  \BibitemOpen
  \bibfield  {author} {\bibinfo {author} {\bibfnamefont {K.}~\bibnamefont {Jakubowski}}, \bibinfo {author} {\bibfnamefont {A.}~\bibnamefont {Chacon}}, \bibinfo {author} {\bibfnamefont {L.~T.}\ \bibnamefont {Tran}}, \bibinfo {author} {\bibfnamefont {A.}~\bibnamefont {Stopic}}, \bibinfo {author} {\bibfnamefont {U.}~\bibnamefont {Garbe}}, \bibinfo {author} {\bibfnamefont {J.}~\bibnamefont {Bevitt}}, \bibinfo {author} {\bibfnamefont {S.}~\bibnamefont {Olsen}}, \bibinfo {author} {\bibfnamefont {D.~R.}\ \bibnamefont {Franklin}}, \bibinfo {author} {\bibfnamefont {A.}~\bibnamefont {Rosenfeld}}, \bibinfo {author} {\bibfnamefont {S.}~\bibnamefont {Guatelli}}, \emph {et~al.},\ }\bibfield  {title} {\bibinfo {title} {A {M}onte {C}arlo model of the {D}ingo thermal neutron imaging beamline},\ }\href@noop {} {\bibfield  {journal} {\bibinfo  {journal} {Sci. Rep.}\ }\textbf {\bibinfo {volume} {13}},\ \bibinfo {pages} {17415} (\bibinfo {year} {2023})}\BibitemShut {NoStop}%
\bibitem [{Video: twoUpTwoDown()}]{suppVideoTwoUpTwoDown}%
  \BibitemOpen
  Video: twoUpTwoDown,\ \href@noop {} {}\bibinfo {note} {See Supplemental Material at [URL will be inserted by publisher] for a video showing the pattern by pattern construction of the {\it twoUpTwoDown} target image (two white squares and two black squares on a grey background, see Fig. \ref{fig:targetImages}b). The process uses 427 total radiographic exposures of 175 unique patterns (or mask positions) and results in the $2 mm \times 2mm$ ghost projection shown in Fig. \ref{fig:gpImages}b.}\BibitemShut {Stop}%
\bibitem [{Video: twoSquares()}]{suppVideoTwoSquares}%
  \BibitemOpen
  Video: twoSquares,\ \href@noop {} {}\bibinfo {note} {See Supplemental Material at [URL will be inserted by publisher] for a video showing the pattern by pattern construction of the {\it twoSquares} target image (two white squares on a black background, see Fig. \ref{fig:targetImages}a). The process uses 550 total radiographic exposures of 179 unique patterns (or mask positions) and results in the $2 mm \times 2mm$ ghost projection shown in Fig. \ref{fig:gpImages}a.}\BibitemShut {Stop}%
\bibitem [{Video: twoSquaresNeg()}]{suppVideoTwoSquaresNeg}%
  \BibitemOpen
  Video: twoSquaresNeg,\ \href@noop {} {}\bibinfo {note} {See Supplemental Material at [URL will be inserted by publisher] for a video showing the pattern by pattern construction of the {\it twoSquaresNeg} target image (two black squares on a white background, see Fig. \ref{fig:targetImages}d). The process uses 385 total radiographic exposures of 167 unique patterns (or mask positions) and results in the $2 mm \times 2mm$ ghost projection shown in Fig. \ref{fig:gpImages}d.}\BibitemShut {Stop}%
\bibitem [{Video: twoUpTwoDownNeg()}]{suppVideoTwoUpTwoDownNeg}%
  \BibitemOpen
  Video: twoUpTwoDownNeg,\ \href@noop {} {}\bibinfo {note} {See Supplemental Material at [URL will be inserted by publisher] for a video showing the pattern by pattern construction of the {\it twoUpTwoDownNeg} target image (two black squares and two white squares on a grey background, see Fig. \ref{fig:targetImages}e). The process uses 541 total radiographic exposures of 181 unique patterns (or mask positions) and results in the $2 mm \times 2mm$ ghost projection shown in Fig. \ref{fig:gpImages}e.}\BibitemShut {Stop}%
\bibitem [{Video: dingoPaw()}]{suppVideoDingoPaw}%
  \BibitemOpen
  Video: dingoPaw,\ \href@noop {} {}\bibinfo {note} {See Supplemental Material at [URL will be inserted by publisher] for a video showing the pattern by pattern construction of the {\it dingoPaw} target image (white dingo paw print on a black background, see Fig. \ref{fig:targetImages}c). The process uses 162 total radiographic exposures of 200 unique patterns (or mask positions) and results in the $2 mm \times 2mm$ ghost projection shown in Fig. \ref{fig:gpImages}c.}\BibitemShut {Stop}%
\bibitem [{Video: dingoSil()}]{suppVideoDingoSil}%
  \BibitemOpen
  Video: dingoSil,\ \href@noop {} {}\bibinfo {note} {See Supplemental Material at [URL will be inserted by publisher] for a video showing the pattern by pattern construction of the {\it DingoSil} target image (white dingo silhouette on a black background, see Fig. \ref{fig:targetImages}f). The process uses 546 total radiographic exposures of 186 unique patterns (or mask positions) and results in the $2 mm \times 2mm$ ghost projection shown in Fig. \ref{fig:gpImages}f.}\BibitemShut {Stop}%
\bibitem [{\citenamefont {Gorban}\ \emph {et~al.}(2016)\citenamefont {Gorban}, \citenamefont {Tyukin}, \citenamefont {Prokhorov},\ and\ \citenamefont {Sofeikov}}]{Gorban2016}%
  \BibitemOpen
  \bibfield  {author} {\bibinfo {author} {\bibfnamefont {A.~N.}\ \bibnamefont {Gorban}}, \bibinfo {author} {\bibfnamefont {I.~Y.}\ \bibnamefont {Tyukin}}, \bibinfo {author} {\bibfnamefont {D.~V.}\ \bibnamefont {Prokhorov}},\ and\ \bibinfo {author} {\bibfnamefont {K.~I.}\ \bibnamefont {Sofeikov}},\ }\bibfield  {title} {\bibinfo {title} {Approximation with random bases: {P}ro et contra},\ }\href@noop {} {\bibfield  {journal} {\bibinfo  {journal} {Inf. Sci.}\ }\textbf {\bibinfo {volume} {364--365}},\ \bibinfo {pages} {129} (\bibinfo {year} {2016})}\BibitemShut {NoStop}%
\bibitem [{\citenamefont {Gureyev}\ \emph {et~al.}(2018)\citenamefont {Gureyev}, \citenamefont {Paganin}, \citenamefont {Kozlov}, \citenamefont {Nesterets},\ and\ \citenamefont {Quiney}}]{Gureyev2018}%
  \BibitemOpen
  \bibfield  {author} {\bibinfo {author} {\bibfnamefont {T.~E.}\ \bibnamefont {Gureyev}}, \bibinfo {author} {\bibfnamefont {D.~M.}\ \bibnamefont {Paganin}}, \bibinfo {author} {\bibfnamefont {A.}~\bibnamefont {Kozlov}}, \bibinfo {author} {\bibfnamefont {Y.~I.}\ \bibnamefont {Nesterets}},\ and\ \bibinfo {author} {\bibfnamefont {H.~M.}\ \bibnamefont {Quiney}},\ }\bibfield  {title} {\bibinfo {title} {Complementary aspects of spatial resolution and signal-to-noise ratio in computational imaging},\ }\href {https://doi.org/10.1103/PhysRevA.97.053819} {\bibfield  {journal} {\bibinfo  {journal} {Phys. Rev. A}\ }\textbf {\bibinfo {volume} {97}},\ \bibinfo {pages} {053819} (\bibinfo {year} {2018})}\BibitemShut {NoStop}%
\bibitem [{\citenamefont {Monro}\ \emph {et~al.}(2024)\citenamefont {Monro}, \citenamefont {Kingston},\ and\ \citenamefont {Paganin}}]{Monro2024}%
  \BibitemOpen
  \bibfield  {author} {\bibinfo {author} {\bibfnamefont {J.~A.}\ \bibnamefont {Monro}}, \bibinfo {author} {\bibfnamefont {A.~M.}\ \bibnamefont {Kingston}},\ and\ \bibinfo {author} {\bibfnamefont {D.~M.}\ \bibnamefont {Paganin}},\ }\href@noop {} {\bibinfo {title} {Ghost projection via focal-field diffraction catastrophes}} (\bibinfo {year} {2024}),\ \Eprint {https://arxiv.org/abs/2411.19053} {arXiv:2411.19053} \BibitemShut {NoStop}%
\bibitem [{\citenamefont {Cand\`{e}s}\ and\ \citenamefont {Tao}(2006)}]{CandesTao2006}%
  \BibitemOpen
  \bibfield  {author} {\bibinfo {author} {\bibfnamefont {E.~J.}\ \bibnamefont {Cand\`{e}s}}\ and\ \bibinfo {author} {\bibfnamefont {T.}~\bibnamefont {Tao}},\ }\bibfield  {title} {\bibinfo {title} {Near-optimal signal recovery from random projections: {U}niversal encoding strategies?},\ }\href@noop {} {\bibfield  {journal} {\bibinfo  {journal} {IEEE Trans. Inf. Theory}\ }\textbf {\bibinfo {volume} {52}},\ \bibinfo {pages} {5406} (\bibinfo {year} {2006})}\BibitemShut {NoStop}%
\bibitem [{\citenamefont {Donoho}(2006)}]{Donoho2006}%
  \BibitemOpen
  \bibfield  {author} {\bibinfo {author} {\bibfnamefont {D.~L.}\ \bibnamefont {Donoho}},\ }\bibfield  {title} {\bibinfo {title} {Compressed sensing},\ }\href {https://doi.org/10.1109/TIT.2006.871582} {\bibfield  {journal} {\bibinfo  {journal} {IEEE Trans. Inf. Theory}\ }\textbf {\bibinfo {volume} {52}},\ \bibinfo {pages} {1289} (\bibinfo {year} {2006})}\BibitemShut {NoStop}%
\bibitem [{\citenamefont {{Rani}}\ \emph {et~al.}(2018)\citenamefont {{Rani}}, \citenamefont {{Dhok}},\ and\ \citenamefont {{Deshmukh}}}]{Rani2017}%
  \BibitemOpen
  \bibfield  {author} {\bibinfo {author} {\bibfnamefont {M.}~\bibnamefont {{Rani}}}, \bibinfo {author} {\bibfnamefont {S.~B.}\ \bibnamefont {{Dhok}}},\ and\ \bibinfo {author} {\bibfnamefont {R.~B.}\ \bibnamefont {{Deshmukh}}},\ }\bibfield  {title} {\bibinfo {title} {A systematic review of compressive sensing: Concepts, implementations and applications},\ }\href@noop {} {\bibfield  {journal} {\bibinfo  {journal} {IEEE Access}\ }\textbf {\bibinfo {volume} {6}},\ \bibinfo {pages} {4875} (\bibinfo {year} {2018})}\BibitemShut {NoStop}%
\bibitem [{\citenamefont {Metcalf}\ and\ \citenamefont {{van der Straten}}(1999)}]{AtomBeamBook}%
  \BibitemOpen
  \bibfield  {author} {\bibinfo {author} {\bibfnamefont {H.~J.}\ \bibnamefont {Metcalf}}\ and\ \bibinfo {author} {\bibfnamefont {P.}~\bibnamefont {{van der Straten}}},\ }\href@noop {} {\emph {\bibinfo {title} {Laser Cooling and Trapping}}}\ (\bibinfo  {publisher} {Springer, New York},\ \bibinfo {year} {1999})\BibitemShut {NoStop}%
\bibitem [{\citenamefont {Robins}\ \emph {et~al.}(2013)\citenamefont {Robins}, \citenamefont {Altin}, \citenamefont {Debs},\ and\ \citenamefont {Close}}]{AtomLasers2013}%
  \BibitemOpen
  \bibfield  {author} {\bibinfo {author} {\bibfnamefont {N.~P.}\ \bibnamefont {Robins}}, \bibinfo {author} {\bibfnamefont {P.~A.}\ \bibnamefont {Altin}}, \bibinfo {author} {\bibfnamefont {J.~E.}\ \bibnamefont {Debs}},\ and\ \bibinfo {author} {\bibfnamefont {J.~D.}\ \bibnamefont {Close}},\ }\bibfield  {title} {\bibinfo {title} {Atom lasers: Production, properties and prospects for precision inertial measurement},\ }\href {https://doi.org/https://doi.org/10.1016/j.physrep.2013.03.006} {\bibfield  {journal} {\bibinfo  {journal} {Phys. Rep.}\ }\textbf {\bibinfo {volume} {529}},\ \bibinfo {pages} {265} (\bibinfo {year} {2013})}\BibitemShut {NoStop}%
\bibitem [{\citenamefont {Kapitza}\ and\ \citenamefont {Dirac}(1933)}]{Kapitza_Dirac_1933}%
  \BibitemOpen
  \bibfield  {author} {\bibinfo {author} {\bibfnamefont {P.~L.}\ \bibnamefont {Kapitza}}\ and\ \bibinfo {author} {\bibfnamefont {P.~A.~M.}\ \bibnamefont {Dirac}},\ }\bibfield  {title} {\bibinfo {title} {The reflection of electrons from standing light waves},\ }\href {https://doi.org/10.1017/S0305004100011105} {\bibfield  {journal} {\bibinfo  {journal} {Math. Proc. Camb. Philos. Soc.}\ }\textbf {\bibinfo {volume} {29}},\ \bibinfo {pages} {297} (\bibinfo {year} {1933})}\BibitemShut {NoStop}%
\bibitem [{\citenamefont {McClelland}\ \emph {et~al.}(1993)\citenamefont {McClelland}, \citenamefont {Scholten}, \citenamefont {Palm},\ and\ \citenamefont {Celotta}}]{Mcclelland1993}%
  \BibitemOpen
  \bibfield  {author} {\bibinfo {author} {\bibfnamefont {J.~J.}\ \bibnamefont {McClelland}}, \bibinfo {author} {\bibfnamefont {R.~E.}\ \bibnamefont {Scholten}}, \bibinfo {author} {\bibfnamefont {E.~C.}\ \bibnamefont {Palm}},\ and\ \bibinfo {author} {\bibfnamefont {R.~J.}\ \bibnamefont {Celotta}},\ }\bibfield  {title} {\bibinfo {title} {Laser-focused atomic deposition},\ }\href {https://doi.org/10.1126/science.262.5135.877} {\bibfield  {journal} {\bibinfo  {journal} {Science}\ }\textbf {\bibinfo {volume} {262}},\ \bibinfo {pages} {877} (\bibinfo {year} {1993})}\BibitemShut {NoStop}%
\bibitem [{\citenamefont {Freimund}\ \emph {et~al.}(2001)\citenamefont {Freimund}, \citenamefont {Aflatooni},\ and\ \citenamefont {Batelaan}}]{Freimund2001}%
  \BibitemOpen
  \bibfield  {author} {\bibinfo {author} {\bibfnamefont {D.~L.}\ \bibnamefont {Freimund}}, \bibinfo {author} {\bibfnamefont {K.}~\bibnamefont {Aflatooni}},\ and\ \bibinfo {author} {\bibfnamefont {H.}~\bibnamefont {Batelaan}},\ }\bibfield  {title} {\bibinfo {title} {Observation of the {K}apitza-{D}irac effect},\ }\href@noop {} {\bibfield  {journal} {\bibinfo  {journal} {Nature}\ }\textbf {\bibinfo {volume} {413}},\ \bibinfo {pages} {142} (\bibinfo {year} {2001})}\BibitemShut {NoStop}%
\bibitem [{\citenamefont {Sethna}(2006)}]{SethnaBook}%
  \BibitemOpen
  \bibfield  {author} {\bibinfo {author} {\bibfnamefont {J.~P.}\ \bibnamefont {Sethna}},\ }\href@noop {} {\emph {\bibinfo {title} {Statistical Mechanics: Entropy, Order Parameters and Complexity}}}\ (\bibinfo  {publisher} {Oxford University Press, Oxford},\ \bibinfo {year} {2006})\BibitemShut {NoStop}%
\bibitem [{\citenamefont {Kravtsov}\ and\ \citenamefont {Orlov}(2012)}]{kravtsov2012caustics}%
  \BibitemOpen
  \bibfield  {author} {\bibinfo {author} {\bibfnamefont {Y.~A.}\ \bibnamefont {Kravtsov}}\ and\ \bibinfo {author} {\bibfnamefont {Y.~I.}\ \bibnamefont {Orlov}},\ }\href@noop {} {\emph {\bibinfo {title} {Caustics, Catastrophes and Wave Fields}}}\ (\bibinfo  {publisher} {Springer-Verlag},\ \bibinfo {address} {Berlin},\ \bibinfo {year} {2012})\BibitemShut {NoStop}%
\end{thebibliography}%

\end{document}